\documentclass[pdflatex,sn-mathphys-num]{sn-jnl}

\usepackage{graphicx}%
\usepackage{multirow}%
\usepackage{amsmath,amssymb,amsfonts}%
\usepackage{amsthm}%
\usepackage{mathrsfs}%
\usepackage[title]{appendix}%
\usepackage{xcolor}%
\usepackage{textcomp}%
\usepackage{manyfoot}%
\usepackage{booktabs}%
\usepackage{algorithm}%
\usepackage{algorithmicx}%
\usepackage{algpseudocode}%
\usepackage{listings}%

\theoremstyle{thmstyleone}%
\theoremstyle{thmstyletwo}%

\theoremstyle{thmstylethree}%

\begin{document}
	
	\title[Moderation Effects and Elasticities in Compositional Regression with a Total]{Moderation Effects and Elasticities in Compositional Regression with a Total. Application to Bayesian Spatiotemporal Modelling of All-cause Mortality from Environmental Stressors}
	
	
	\author[1,2]{\fnm{Germ\`a} \sur{Coenders}}\email{germa.coenders@udg.edu}
	
	\author[3]{\fnm{Javier} \sur{Palarea-Albaladejo}}\email{javier.palarea@udg.edu}
	
	\author*[1,2]{\fnm{Marc} \sur{Saez}}\email{marc.saez@udg.edu}
	
	\author[1,2]{\fnm{Maria A.} \sur{Barcel\'o}}\email{antonia.barcelo@udg.edu}

	\affil[1]{\orgdiv{Research Group on Statistics, Econometrics and Health (GRECS)}, \orgname{University of Girona}, \orgaddress{\street{C. Universitat de Girona 10}, \city{Girona}, \postcode{17003}, \country{Spain}}}
	
	\affil[2]{\orgdiv{Centro de Investigaci\'on Biom\'edica en Red de Epidemiolog\'ia 
			y Salud P\'ublica}, \orgname{Instituto de Salud Carlos III}, \orgaddress{\street{Av. Monforte de Lemos 5-7}, \city{Madrid}, \postcode{28029}, \country{Spain}}}
	
	\affil[3]{\orgdiv{Dept. of Computer Science, Applied Mathematics and Statistics}, \orgname{University of Girona}, \orgaddress{\street{C. Maria Aur\`elia Capmany 61}, \city{Girona}, \postcode{17003}, \country{Spain}}}

	
	\abstract{Compositional regression models with a real-valued response variable can generally be specified  as log-contrast models subject to a zero-sum constraint on the model coefficients. This formulation emphasises the relative information conveyed in the composition, while the overall total is regarded irrelevant. In this work, such a setting is extended to account not only for total effects, formally defined in a so-called  $\mathcal{T}$-space, but also for moderation or interaction effects. This is applied in the context of complex spatiotemporal data modelling, through an adaptation of the integrated nested Laplace approximation (INLA) method within a Bayesian estimation framework. Particular emphasis is placed on the interpretation of model coefficients and results, both on the original scale of the response variable and in terms of elasticities.
		
		The methodology is demonstrated through a detailed case study investigating the relationship between all-cause mortality and the interaction between extreme temperatures, air pollution composition, and total air pollution in Catalonia, Spain, during the summer of 2022. The results indicate that extreme temperatures are associated with an increased risk of mortality four days after exposure. Additionally, exposure to total air pollution, especially to NO\textsubscript{2}, is linked to elevated mortality risk regardless of temperature. In contrast, particulate matter is associated to increased mortality only when exposure occurs on days of extreme heat.}

	
	
	
	\maketitle

	\section{Introduction}\label{s_intro}
	
	Compositional data arise in various scientific disciplines where multivariate non-negative observations referring to parts of a whole are collected. 
	In practice, they might be originally expressed in any relative unit of measurement (e.g. parts per million, ppm; weight percentages, wt\%; milligrammes per litre, mg/L; and so on), although they are often rescaled (\emph{closed}) by dividing each part by their total sum. Thus, they are equivalently re-expressed in proportions or percentages. This is commonly the case when referring to, for instance, nutritional compositions, chemical concentrations, unemployment rates across regions, and the like. Note that, in doing so, the researcher implicitly assumes that the total is irrelevant for the scientific question and, formally, the data are projected onto a simplex as sample space \citep{Ait1982}. Importantly, the relevant relative information contained in the data remains the same regardless of the scale and, hence, the results of any statistical analysis should be compatible. In air pollution studies, the compositional approach enables taking the full pollution mix and its intrinsic interdependences into account, rather than focusing on a specific pollutant \citep{sanchez2020spatio}. A recent overview of the state of the art in compositional data analysis since Aitchison's early work can be found in \citep{coenders202340}.
	
	It has been thoroughly shown that applying conventional statistical methods directly to compositional data can produce misleading or spurious results \citep[see e.g][as general references]{aitchison1986statistical, pawlowsky2015coda, filzmoser2018applied}. Instead, the mainstream approach involves mapping them onto the real space through the definition of sensible log-ratios between the parts, which focuses the data analysis solely on the relative information. By using log-ratio transformations or coordinates, statistical analysis can, with some caveats, be equivalently conducted through standard methods for real-valued data.
	
	The so-called additive log-ratio (alr) and orthonormal/isometric log-ratio (olr/ilr) coordinates have been commonly used in regression analysis with compositional covariates \citep{aitchison1984log,coenders2020interpretations,BFHTT2021}. Working with log-ratios guarantees desirable properties, such as the irrelevance of both the scale and the number of parts forming the composition. It also prevents technical issues with closed compositional data, such as singularity of covariance matrices and perfect multicollinearity in regression analysis. Note that these log-ratio representations are particular cases of a log-contrast, i.e. a linear combination of the log-transformed parts of the composition where the coefficients are constrained to add up to zero, so that the expression remains scale invariant. As detailed later on, log-contrasts will be used as the basis to formulate our modelling approach.
	
	However, there are situations where the practitioner is actually interested in the potential effect of the total value of the composition in their modelling, so long as this is not constant (i.e. the data are not closed). That is, not only the relative, but also the absolute information is relevant for the scientific question therein. This is clearly the case in air pollution studies, where total pollution levels should not be ignored \citep{mota2022compositional}. 
	Combining both relative and absolute information coherently within a statistical model draws on the theory of $\mathcal{T}$-spaces \citep{pawlowsky2015tools}. In regression analysis, this allows for the separation of the effects of relative and absolute changes in the composition. It is important to note that simply applying a logarithm transformation of the compositional variables, as often done in environmental sciences (e.g. when analysing pollutant concentrations), fails to achieve such separation, since log-transformed values inherently mix both types of information \citep{coenders2017relative}.
	
	In this work, compositional regression analysis with a total is extended to encompass potential moderation or interaction effects. Briefly, moderation analysis allows exploring how a certain variable influences the strength or direction of the relationship between explanatory and response variables, which in regression implies considering an interaction term between them \citep{jaccard2003interaction}. With a focus on interpretability, a novel parameterisation is formulated which enables the discussion of results in terms of elasticities. This can, in fact, be readily integrated into wider families of models and accommodate both frequentist and Bayesian inference approaches. In particular, the required zero-sum constraint on the model coefficients to account for compositionality is embedded here for the first time into the integrated nested Laplace approximation (INLA) method for Bayesian inference.
	
	In the following, Section \ref{s_logcontrast} presents the compositional regression model with compositional covariates and a total in terms of log-contrasts and some issues when fitting such models with Bayesian inference. Section \ref{s_moderation} elaborates on its extension to consider moderation effects and the interpretation of the associated model coefficients. To illustrate the use, in Section \ref{s_illustration} the method is embedded in Bayesian spatiotemporal modelling, using a novel adaptation of the INLA method to deal with the compositional aspects. This model is applied to investigate the relationship between all-cause mortality and the interaction between extreme temperatures, air pollution composition, and total air pollution in the region of Catalonia, Spain, during the summer of 2022. Finally, Section \ref{s_disc} concludes with some final remarks. 
	
	\section{Compositional linear regression model with real-valued response and compositional covariates}\label{s_logcontrast}
	
	Let us consider a non-closed composition ${\bf{x}}$ comprising $D$ strictly positive parts, that is
	
	\begin{equation}
		\label{e_comp}
		{\bf{x}} = \left( {{x_1},\;{x_2},\; \ldots ,\;{x_D}} \right) \in {\mathbb{R}^{D + }},\;{\rm{with}}\;{x_j} > 0.
	\end{equation}
	
	The purpose is to model the association between ${\bf{x}}$, in the role of explanatory or predictor variable, and a real-valued variable $y$ in the role of dependent or response variable. The most fundamental form of a linear regression model in this setting is the so-called log-contrast model, as originally formulated in \cite{aitchison1984log}. However, following \cite{muller2018interpretation} and \cite{coenders2020interpretations}, here we specify the log-contrast element of the model in terms of logarithms in base 2 to simplify the interpretation of its coefficients:
	
	\begin{equation}
		\label{e_logcontrast}
		\begin{array}{l}
			y = {\beta _0} + {\beta _1}{\log _2}\left( {{x_1}} \right) + {\beta _2}{\log _2}\left( {{x_2}} \right) +  \cdots  + {\beta _D}{\log _2}\left( {{x_D}} \right) + \varepsilon ,\quad \\
			{\rm{with}}\;\sum\limits_{j = 1}^D {{\beta _j}}  = 0 \mbox{ and } \varepsilon\sim N(0,\sigma^2).
		\end{array}
	\end{equation}
	
	The constraint $\sum\limits_{j = 1}^D {{\beta _j}}  = 0$, apart from guaranteeing scale invariance of the log-contrast as noted above, ensures in practice that not all parts can simultaneously increase in relative terms. The $\varepsilon$ corresponds to the normally distributed error term of mean zero and constant variance $\sigma^2$, as commonly included in linear regression models.
	
	Recall that compositionality implies that a part can increase if and only if at least one other part decreases. An increase in parts with positive coefficients $\beta_j$ at the expense of a decrease in parts with negative coefficients $\beta_j$  will be associated with a higher expected value of the response $y$. In more precise terms, due to the use of logarithms in base 2, a coefficient $\beta_j$ is interpreted as the expected increase in $y$ when the ratio of a part $x_j$ to any other part doubles \citep{coenders2020interpretations}.
	
	\subsection{Model fitting}\label{ss_estimation}
	
	In an ordinary frequentist estimation setting, the above model \eqref{e_logcontrast} can be directly fitted by the constrained least squares method, subject to $\sum\limits_{j = 1}^D {{\beta _j}}  = 0$, or by ordinary least squares after the alr transformation is applied on the $D$-part composition {\bf{x}} \citep{aitchison1984log}. That is, using e.g. the last part $x_D$ as reference in the denominator of the log-ratios without loss of generality, the $(D-1)$-dimensional alr-vector is given by $(x_1/x_D,x_2/x_D,\ldots,x_{D-1}/x_D)$, and the model is then specified as
	\begin{equation}
		\label{e_alr}
		\begin{array}{l}
			y = {\beta _0} + {\beta _1}{\log _2}\left( {\frac{{{x_1}}}{{{x_D}}}} \right) + {\beta _2}{\log _2}\left( {\frac{{{x_2}}}{{{x_D}}}} \right) +  \cdots  + {\beta _{D - 1}}{\log _2}\left( {\frac{{{x_{D - 1}}}}{{{x_D}}}} \right) + \varepsilon ,\;\\
			{\rm{with}}\;{\beta _D} =  - {\beta _1} - {\beta _2} -  \cdots  - {\beta _{D - 1}} \mbox{ and } \varepsilon\sim N(0,\sigma^2).
		\end{array}
	\end{equation}
	
	Note that using the least squares method, it does not matter which part is chosen as denominator for the alr representation. Thus, the same results would be obtained using, say, the part $x_1$ as reference instead of $x_D$, leading to the model specification
	
	\begin{equation}
		\label{e_alr2}
		\begin{array}{l}
			y = {\beta _0} + {\beta _2}{\log _2}\left( {\frac{{{x_2}}}{{{x_1}}}} \right) + {\beta _3}{\log _2}\left( {\frac{{{x_3}}}{{{x_1}}}} \right) +  \cdots  + {\beta _D}{\log _2}\left( {\frac{{{x_D}}}{{{x_1}}}} \right) + \varepsilon ,\;\\
			{\rm{with}}\;{\beta _1} =  - {\beta _2} - {\beta _3} -  \cdots  - {\beta _D} \mbox{ and } \varepsilon\sim N(0,\sigma^2).
		\end{array}
	\end{equation}
	
	However, it has been shown that this will not necessarily hold for other estimation methods. This is the case of Bayesian inference using INLA 
	\citep{gomez2020bayesian,sanchez2021compositional}
	or Markov  chain Monte Carlo (MCMC) estimation \citep{le2024bayesian,sanchez2020spatio}. Here, the prior distribution interferes with the log-ratio representation, making the model specifications (\ref{e_alr}) and (\ref{e_alr2}) no longer equivalent as it should be. 
	
	Within a Bayesian framework, the fitting of model (\ref{e_logcontrast}) requires incorporating the compositional constraint $\sum\limits_{j = 1}^D {{\beta _j}} = 0$ into the specification of the prior distributions. 
	Following 
	\cite{zhang2025bayesian}, \cite{zhang2024bayesianordinal} and \cite{zhang2024bayesianglm}, this can be achieved using MCMC methods by implementing a \emph{soft} constraint of the form
	\begin{equation}
		\label{e_softconstraint}
		\sum\limits_{j = 1}^D {{\beta _j}}  \sim N\left( {0,\;0.001D} \right).
	\end{equation}
	
	Or, alternatively, it can be done by assuming the \emph{hard} constraint $\sum\limits_{j = 1}^D {{\beta _j}}  = 0$ as in \cite{scott2023bayesian}. The latter implies that the multivariate priors for the vector of model coefficients $({\beta _1}, \dots, {\beta _D})$  will have a singular covariance matrix. Note that none of these approaches has been considered for the INLA estimation method so far.
	
	Importantly, including the constraint, either in its soft or hard variant, ensures the reproducibility of results and, at the same time, avoids the need to set an arbitrary reference compositional part to apply the alr transformation. This is also compatible with the principle of permutation invariance in compositional data analysis \citep{aitchison1986statistical}, by which the statistical results should be invariant to permutations of the parts of the composition.
	
	This is the approach we adhere to in the current work, particularly through the novel implementation of the soft constraint in the INLA method.  This is done on the R system for statistical computing \citep{r2025r} through a new function called  \texttt{A.local} in the package R-INLA \citep{inlaproject2023}. The \texttt{A.local} function allows users to impose local constraints on the linear predictor. Previously, if a linear predictor contained more than one element of the same model component, it required a global redefinition through complicated design matrices, which also involved irrelevant parts of the model. With \texttt{A.local}, this is avoided in most cases by efficiently applying constraints directly to specific subsets of the model predictors.
	
	\subsection{Considering the total effect}\label{ss_total}
	
	As previously discussed, there are cases where, in addition to the potential effect of relative changes among parts of the composition, their actual absolute values may also be contributing to the variation in the response variable \citep{coenders2017relative}. To account for this, a term capturing the total magnitude of the composition can be included into the regression model. A coherent way to do this, while respecting the geometric properties of the sample space of compositions, is provided by the theory of $\mathcal{T}$-spaces \citep{pawlowsky2015tools}. A few different formulations have been discussed to summarise such total information, with the so-called multiplicative total being preferred according to \cite{pawlowsky2015tools}. This is defined as
	
	\begin{equation}
		\label{e_multtotal}
		\sqrt D \log \left( {\sqrt[D]{{{x_1}{x_2} \cdots {x_D}}}} \right) = \frac{1}{{\sqrt D }}\left( {\log ({x_1}) + \log ({x_2}) +  \cdots  + \log ({x_D})} \right).
	\end{equation}
	
	However, for estimation and interpretation purposes within a linear modelling framework using INLA, \cite{mota2024airpollution} simplified it to
	
	\begin{equation}
		\label{e_multtotal2}
		\begin{array}{l}
			t = D{\log _2}\left( {\sqrt[D]{{{x_1}{x_2} \cdots {x_D}}}} \right) = {\log _2}\left( {{x_1}{x_2} \cdots {x_D}} \right) = \\
			\left( {{{\log }_2}({x_1}) + {{\log }_2}({x_2}) +  \cdots  + {{\log }_2}({x_D})} \right).
		\end{array}
	\end{equation}
	
	Hence, a log-contrast model \eqref{e_logcontrast} extended to include the information about the total \eqref{e_modeltotal2} is written as
	\begin{equation}
		\label{e_modeltotal}
		\begin{array}{l}
			y = {\beta _0} + {\beta _1}{\log _2}\left( {{x_1}} \right) + {\beta _2}{\log _2}\left( {{x_2}} \right) +  \cdots  + {\beta _D}{\log _2}\left( {{x_D}} \right) + {\beta ^{\left( t \right)}}t + \varepsilon ,\quad \\
			{\rm{with}}\;\sum\limits_{j = 1}^D {{\beta _j}}  = 0 \mbox{ and } \varepsilon\sim N(0,\sigma^2).
		\end{array}
	\end{equation}
	
	The new coefficient ${\beta ^{\left( t \right)}}$ can be interpreted as $1/D$ times the effect on the expected value of the response $y$ when the absolute values of all parts are simultaneously doubled, while preserving the relative structure encoded in their log-ratios. Moreover, a coefficient $\beta_j$, for $j=1,\ldots, D$, is interpreted as the expected increase in the response $y$ when the ratio of the part $x_j$ to any other part doubles, while the total of the composition is kept constant. This means that the parts $x_1$, $x_2,\ldots, x_{j-1}, x_{j+1},\ldots, x_D$ decrease by a common factor to compensate for the increase in $x_j$, thereby keeping the total $t$, as defined in (\ref{e_multtotal2}), constant \citep{coenders2017relative}. 
	
	
	In a classical inference setting, the significance of the effect of the compositional (relative information) term is tested by defining the joint null hypothesis ${{\rm{H}}_{\rm{0}}}:{\beta _1} = {\beta _2} =  \cdots  = {\beta _D} = 0$. Moreover, the significance of the effect of the total (absolute information) term is tested by defining the null hypothesis ${{\rm{H}}_{\rm{0}}}:{\beta ^{(t)}} = 0$, and then using $F$ and Student's $t$ statistics respectively as ordinarily.
	
	Following \cite{mota2024airpollution}, and with the aim of making the decomposition between compositional (zero-sum) and total (constant) effects more evident, model \eqref{e_modeltotal} can be reformulated as
	\begin{equation}
		\label{e_modeltotal2}
		\begin{array}{l}
			y = {\beta _0} + \left( {{\beta _1} + {\beta ^{\left( t \right)}}} \right){\log _2}\left( {{x_1}} \right) + \left( {{\beta _2} + {\beta ^{\left( t \right)}}} \right){\log _2}\left( {{x_2}} \right) +  \cdots  + \left( {{\beta _D} + {\beta ^{\left( t \right)}}} \right){\log _2}\left( {{x_D}} \right) + \varepsilon ,\quad \\
			{\rm{with}}\;\sum\limits_{j = 1}^D {{\beta _j}}  = 0 \mbox{ and } \varepsilon\sim N(0,\sigma^2).
		\end{array}
	\end{equation}
	
	In this case, a coefficient $({\beta _j} + {\beta ^{\left( t \right)}})$ is interpreted as the effect of multiplying $x_j$ by 2 while keeping the absolute values of the parts $x_1$, $x_2,..., x_{j-1}, x_{j+1},..., x_D$ constant. Furthermore, if $y$ is log-transformed, the model becomes
	\begin{equation}
		\label{e_modeltotal3}
		\begin{array}{l}
			{\log _2}\left( y \right) = {\beta _0} + \left( {{\beta _1} + {\beta ^{\left( t \right)}}} \right){\log _2}\left( {{x_1}} \right) + \left( {{\beta _2} + {\beta ^{\left( t \right)}}} \right){\log _2}\left( {{x_2}} \right) +  \cdots  + \left( {{\beta _D} + {\beta ^{\left( t \right)}}} \right){\log _2}\left( {{x_D}} \right)\\
			+ \varepsilon ,\quad {\rm{with}}\;\sum\limits_{j = 1}^D {{\beta _j}}  = 0 \mbox{ and } \varepsilon\sim N(0,\sigma^2),
		\end{array}
	\end{equation}
	
	\noindent and the coefficients are readily interpreted as elasticities.
	
	Alternative formulations of compositional regression models have been proposed to enable interpretation in terms of elasticities \citep{dargel2024link, morais2021impact}. However, these approaches involve complex functions of the model parameters. Contrarily, in our proposed model \eqref{e_modeltotal3}, the sum of coefficients $({\beta _j} + {\beta ^{\left( t \right)}})$ is directly interpretable as an elasticity. Thus, if a part $x_j$ has a small relative increase (say by 1\%), while keeping the remaining parts unaltered, the response variable $y$ will increase by $({\beta _j} + {\beta ^{\left( t \right)}})$\%.
	
	This elasticity can be further decomposed into its compositional and total contributions. On the one hand, if the ratio of a part $x_j$ to any other part increases by 1\%, while keeping the total constant, then $y$ will increase by $\beta_j$\%. On the other hand, if all parts increase simultaneously by $1/D$\%, then $y$ will increase by ${\beta ^{\left( t \right)}}$\%.
	
	Finally, note that the base of the logarithm used in \eqref{e_modeltotal3} does not affect the interpretation, as long as it is applied consistently to both the response and explanatory sides of the model.
	
	\section{Adding a moderation effect}\label{s_moderation}
	
	\subsection{Formulation in the original scale of the response variable}
	\label{s_moderation1}
	
	Building on model \eqref{e_modeltotal}, we introduce the extension to incorporate a moderation or interaction effect. This involves the composition, the total, and a new moderating variable $z$. The variable $z$ may be defined either as a dummy categorical variable, coded $\{0, 1\}$, or as a numeric variable. In the latter case, we assume $z$ is mean-centred to aid interpretation. Similarly, $\log _2\left( {{x_1}} \right), \ldots, \log _2\left( {{x_D}} \right)$ and $t$ are also centred around their respective means. The full model can thus be written as
	\begin{equation}
		\label{e_moderation1}
		\begin{array}{l}
			y = {\beta _0} + {\beta _1}{\log _2}\left( {{x_1}} \right) + {\beta _2}{\log _2}\left( {{x_2}} \right) +  \cdots  + {\beta _D}{\log _2}\left( {{x_D}} \right) + {\beta ^{(z)}}z + \\
			+ \beta _1^{(i)}z{\log _2}\left( {{x_1}} \right) + \beta _2^{(i)}z{\log _2}\left( {{x_2}} \right) +  \cdots  + \beta _D^{(i)}z{\log _2}\left( {{x_D}} \right) + \\
			+ {\beta ^{(t)}}t + {\beta ^{(t,i)}}tz + \varepsilon ,\quad \\
			{\rm{with}}\;\sum\limits_{j = 1}^D {{\beta _j}}  = 0\;{\rm{,}}\;\sum\limits_{j = 1}^D {\beta _j^{(i)}}  = 0 \mbox{ and } \varepsilon\sim N(0,\sigma^2).
		\end{array}
	\end{equation}
	
	The following points summarise the interpretation of the model coefficients:
	
	\begin{itemize}
		
		\item The coefficients $\beta_j$, ${\beta ^{\left( t \right)}}$, and ${\beta ^{\left( z \right)}}$ represent the main effects, each corresponding to the scenario where the other variables are held at zero. Specifically, the coefficient $\beta_j$ quantifies the expected change in $y$ when the ratio of $x_j$ to any other part doubles, assuming that the total remains constant and $z=0$ (i.e. the variable $z$ is held at its mean value if numeric or at the reference level 0 if binary). The coefficient ${\beta ^{\left( t \right)}}$ captures the effect (scaled by $1/D$) of simultaneously doubling the absolute values of all parts of the composition, while $z=0$ is assumed. Thanks to the mean-centring, the coefficient ${\beta ^{\left( z \right)}}$ represents the expected change in $y$ associated with a one-unit increase in $z$ (or a change from level 0 to 1 in the binary case), when the terms $\log _2\left( {{x_1}} \right), \ldots, \log _2\left( {{x_D}} \right)$ are held at zero (i.e. at their respective mean values).
		
		
		\item The coefficients ${\beta ^{\left( i \right)}}$ account for the moderating or interaction effects, i.e. the change in the compositional and total effects when $z=1$ (meaning that the variable $z$ is one unit above its mean value if numeric or at level 1 if binary). In other words, when $z=0$, the log-contrast coefficients are ${\beta _1},{\beta _2}, \cdots ,{\beta _D}$; whereas when $z=1$, the log-contrast coefficients become $({\beta _1} + \beta _1^{(i)}),({\beta _2} + \beta _2^{(i)}), \ldots ,({\beta _D} + \beta _D^{(i)})$.
		
	\end{itemize}
	
	Statistical testing of the significance of model coefficient can be conducted as usually. Thus, the compositional main effect can be tested by setting the null hypothesis ${{\rm{H}}_{\rm{0}}}:{\beta _1} = {\beta _2} =  \cdots  = {\beta _D} = 0$, with the caveat that it only applies to the case of $z=0$. Similarly, the significance of the total main effect can be tested by setting ${{\rm{H}}_{\rm{0}}}:{\beta ^{(t)}} = 0$. Moreover, the assessment of the $z$ main effect, via ${{\rm{H}}_{\rm{0}}}:{\beta ^{(z)}} = 0$, considers the composition set at its mean value.
	
	The significance of the interaction between the composition and the moderating variable $z$ can be tested by the null hypothesis ${{\rm{H}}_0}:\beta _1^{(i)} = \beta _2^{(i)} =  \cdots  = \beta _D^{(i)} = 0$. The coefficient $\beta ^{(t,i)}$ also refers to an interaction effect, in this case indicating changes in the total effect when $z$ either raises one unit over the mean (numeric case) or shifts from level 0 to 1 (binary case). That is, when $z=0$ the total effect is just ${\beta ^{\left( t \right)}}$, but for $z=1$ it includes the term of the interaction with $z$ and is given by $(\beta ^{(t)} + \beta ^{(t,i)})$. The significance of this \emph{moderated} total effect can be tested through the null hypothesis ${{\rm{H}}_0}:{\beta ^{(t,i)}} = 0$.
	
	The joint significance of all interaction effects can be assessed by the null hypothesis ${{\rm{H}}_0}:\beta _1^{(i)} = \beta _2^{(i)} =  \cdots  = \beta _D^{(i)} = {\beta ^{(t,i)}} = 0$. Note that these interaction effects are often of substantial practical interest. For instance, the impact of a relative increase in a specific part $x_j$ on the response $y$ may vary depending on the level of $z$. This is governed by the sign of $\beta _j^{(i)}$: a positive value indicates a gain in influence with increasing values of $z$, while a negative value indicates a loss. Similarly, the impact of an absolute increase in all parts of the composition on $y$ with higher values of $z$ will be conditioned by the sign of $\beta _{}^{(t,i)}$, being strengthened or weakened in the case of positive or negative signs respectively.
	
	\subsection{Formulation in terms of elasticities}
	\label{s_moderation2}
	
	As discussed in Section \ref{s_logcontrast}, expressing the response variable $y$ in logarithmic scale in the model allows a moderation effect to be interpreted in terms of changes in elasticities. Thus, specifying the model in the form
	
	\begin{equation}
		\label{e_moderation2}
		\begin{array}{l}
			{\log _2}\left( y \right) = {\beta _0} + {\beta _1}{\log _2}\left( {{x_1}} \right) + {\beta _2}{\log _2}\left( {{x_2}} \right) +  \cdots  + {\beta _D}{\log _2}\left( {{x_D}} \right) + {\beta ^{(z)}}z + \\
			+ \beta _1^{(i)}z{\log _2}\left( {{x_1}} \right) + \beta _2^{(i)}z{\log _2}\left( {{x_2}} \right) +  \cdots  + \beta _D^{(i)}z{\log _2}\left( {{x_D}} \right) + \\
			+ {\beta ^{(t)}}t + {\beta ^{(t,i)}}tz + \varepsilon ,\quad \\
			{\rm{with}}\;\sum\limits_{j = 1}^D {{\beta _j}}  = 0\;{\rm{,}}\;\sum\limits_{j = 1}^D {\beta _j^{(i)}} = 0 \mbox{ and } \varepsilon\sim N(0,\sigma^2),
		\end{array}
	\end{equation}
	
	leads to the following interpretations:
	
	\begin{itemize}
		
		\item Regarding main effects, the sum $({\beta _j} + {\beta ^{\left( t \right)}})$ is directly interpretable as an elasticity. Thus, a small relative increase (e.g. by 1\%) in $x_j$, while keeping the remaining parts unaltered, leads to an increase of $\left({\beta _j} + {\beta ^{\left( t \right)}}\right)$\% in the response $y$. Note that this is so only when $z=0$ (i.e. with $z$ at its mean value if numeric or at the reference level 0 if binary). Furthermore, these elasticities can be decomposed into total and compositional effects as follows. When all parts of the composition increase simultaneously by $1/D$\%, then the response $y$ increases by ${\beta ^{\left( t \right)}}$\% while holding $z=0$. When the ratio of $x_j$ to any other part increases by 1\%, while keeping the total constant, the response $y$ increases by $\beta_j$\%, with $z=0$. 
		
		\item The coefficient $\beta ^{(z)}$ represents a semi-elasticity. When $z$ increases by one unit (above its mean value in the numerical case or from level 0 to 1 in the binary case), the response $y$ increases by a factor of $2^{\beta ^{(z)}}$, for $\log _2\left( {{x_1}} \right), \ldots, \log _2\left( {{x_D}} \right)$ held at value zero (i.e. at the mean composition).
		
		\item As to the moderation or interaction effects, the coefficients $\beta _j^{(i)}$ and  $\beta ^{(t,i)}$ represent the change in the compositional and total effects for the case of $z=1$ (i.e. when $z$ is one unit above its mean value if numeric or at level 1 if binary). If a certain part $x_j$ increases by 1\%, while the other parts remain constant, the response $y$ increases by $({\beta _j} + \beta _{}^{\left( t \right)} + \beta _j^{\left( i \right)} + \beta _{}^{\left( {t,i} \right)})$\%. As noted before, these elasticities can be further decomposed into total and compositional effects. Thus, if all the parts increase simultaneously by $1/D$\%, then the response $y$ increases by $({\beta ^{\left( t \right)}} + {\beta ^{\left( {t,i} \right)}})$\%. If the ratio of $x_j$ to any other part increases by 1\%, the response $y$ increases by $({\beta _j} + \beta _j^{\left( i \right)})$\%.
		
	\end{itemize}
	
	Any hypothesis testing on the coefficients of model \eqref{e_moderation2} would be conducted in the same manner as for the model expressed on the original scale of the response variable \eqref{e_moderation1}, as described in Section \ref{s_moderation1}. We emphasise once again that the base of the logarithm does not affect the interpretation of the compositional and total effects, as long as the base is applied consistently to both the response and explanatory sides of the model. Only the interpretation of the main effect of $z$ will change, referring to an increase in $y$ by a factor $b^{\beta^{(z)}}$, where $b$ stands for the base of the logarithm.
	
	Finally, note that all the above, elaborated for the ordinary linear regression model, can be readily extended to generalised linear models (GLMs) with non-normal response variables \citep{coenders2017relative}. The case study developed in the following section provides an illustration of such extension using a zero-inflated negative binomial regression model.
	
	\section{Bayesian spatiotemporal modelling of all-cause mortality from environmental stressors}\label{s_illustration}
	
	\subsection{Background}
	\label{s_backbround}
	
	In environmental health studies, the compositional data methodology has been used to identify patterns of pollution which can be associated with health risks \citep{sanchez2020spatio,sanchez2022spatially}. In this context, jointly considering relative and absolute information, through their representation in a $\mathcal{T}$-space, makes it possible to study the relative importance of pollutant concentrations along with the impact of the overall pollution level \citep{mota2022compositional,mota2024airpollution}. Furthermore, it is well known that exposure to certain air pollutants can modify the impact of extreme heat on mortality. In particular, it has been found that ozone exposure influences the effects of heatwaves on cardiovascular, respiratory, and overall mortality \citep{alari2023role,analitis2014effects,dear2005effects,du2024exposure,qi2023modification}. The effect of the interaction between ozone concentration and temperature has also  been discussed in relation to other health parameters such as birth weight \citep{Martenies2022using} or sperm quality \citep{Wang2020sperm}. Moreover, exposure to PM\textsubscript{10} and PM\textsubscript{2.5} have also been found to play some moderating role on mortality \citep{analitis2014effects,Landguth2024seasonal,xu2023extreme} .
	
	When working within a spatiotemporal framework, the estimation of a second-order stationary Gaussian field (GF) is known to often suffer from the so-called big $n$ problem. This refers to the high computational burden associated to model fitting with increasing size and variability along both space and time dimensions \citep{lindgren2015bayesian}. Computational burden can be alleviated by representing a GF as a Gaussian Markov random field (GMRF) \citep{rue2009approximate}. This is defined by a precision matrix with a sparse structure, allowing inference to be performed in a computationally efficient way. The GF and GMRF are linked through the stochastic partial differential equations approach \citep{lindgren2011explicit}. This makes it possible to find a GMRF with local neighbourhood and sparse precision matrix (instead of the spatiotemporal covariance function and dense covariance matrix of a GF) which best represents a Mat\'ern field \citep{cameletti2013spatio,lindgren2011explicit}. This often renders a frequentist approach impractical, favouring the use of Bayesian alternatives instead. In this setting, the INLA approach, as used in this work, has been shown to be much faster and computationally efficient than traditional Bayesian methods based on MCMC, producing accurate approximations to posterior distributions, even for very complex models \citep{lindgren2015bayesian}. In this vein, the model specification detailed in \cite{saez2022spatial} offers a fast yet sufficiently accurate solution. Such specification is used as reference for the model presented here.
	
	Our application concerns an ecological small-area study conducted at the level of the 379 Basic Health Areas (ABS) in which the territory of Catalonia is divided. The dataset comprises daily observations from June to September 2022. In the summer of 2022, Spain as a whole experienced an unprecedented series of 41 consecutive days of extreme temperatures due to three successive heatwaves. This broke previous records in 2015 and 2004 by 12 and 21 days respectively. The heatwave in July 2022 hit 44 out of the 52 Spanish provinces, surpassing any previous event in terms of number of provinces affected and registering a temperature anomaly of 3.7$^\circ$C above the monthly average. In addition, the first heatwave (12-18 June) was the second-earliest on record, whereas the second heatwave (30 July to 15 August) was the most extensive and intense observed to date \citep{AEMET}. In Catalonia specifically, the summer of 2022 was the hottest on record, with over two weeks of sustained above-average temperatures in July, ranking among the most persistent heatwaves experienced so far in the region \citep{METEOCAT}.
	
	\subsection{Model formulation}
	\label{s_results}
	
	The use of the methodology introduced above is adapted and illustrated here in the context of Bayesian modelling of environmental spatiotemporal data using the INLA method for parameter estimation. The purpose is to investigate the association between all-cause mortality as response variable and extreme temperatures, air pollution and the interaction between them as explanatory variables.
	
	The mortality data used were obtained from the Spanish National Statistics Institute (INE) and consist of daily all-cause mortality counts for the summer months (June to September) of 2022. They are aggregated at the ecological level for each ABS and are available for the total population.
	
	The meteorological data were provided by the XEMA network (Network of Automatic Meteorological Stations; open data available at \href{https://analisi.transparenciacatalunya.cat/ca/Medi-Ambient/Dades-meteorol-giques-de-la-XEMA/nzvn-apee/about_data}{https://analisi.transparenciacatalunya.cat/ca/Medi-Ambient/Dades-meteorol-giques-de-la-XEMA/nzvn-apee/about\_data}), which comprises 189 automatic weather stations distributed across Catalonia. This excluded 12 stations located at altitudes of 1,500 metres or higher \citep{METEOCATDades}.
	The key variable here is the maximum daily temperature.
	
	Lastly, the pollution data are daily averages of hourly levels of the concentrations of five pollutants: particulate matter (PM\textsubscript{10}), nitrogen dioxide (NO\textsubscript{2}), ozone (O\textsubscript{3}), carbon monoxide (CO) and sulphur dioxide (SO\textsubscript{2}). These were measured for the 95 automatic monitoring stations of the XVPCA network (Catalan Network for Pollution Control and Prevention; open data available at \href{https://analisi.transparenciacatalunya.cat/Medi-Ambient/Qualitat-de-l-aire-als-punts-de-mesurament-autom-t/tasf-thgu/about\_data}{https://analisi.transparenciacatalunya.cat/Medi-Ambient/Qualitat-de-l-aire-als-punts-de-mesurament-autom-t/tasf-thgu/about\_data}) \citep{XVPCA}. Note that for ABSs lacking a nearby weather or pollution monitoring station in their territory, the corresponding meteorological or pollutant concentrations were estimated using a spatiotemporal predictive model following the approach in \citep{mota2022compositional,saez2022spatial}.
	
	
	Building on the set-up of model \eqref{e_moderation2}, we adapted it to account for the fact that the response variable, all-cause mortality, arises from a count process and that random effects are included. Thus, model selection was confined to the family of generalised linear mixed models (GLMMs) within the Bayesian framework provided by the R-INLA package \citep{rue2009approximate,rue2017bayesian} used in experimental mode \citep{van2023new}. These included Poisson and negative binomial distribution specifications for the response variable, with the latter accounting for data overdispersion. Moreover, since a number of ABSs did not report any deaths over several days, we also considered their zero-inflated variants. Based on the Watanabe-Akaike Information Criterion (WAIC) \citep{watanabe2010asymptotic}, the best fit to the data (lowest WAIC) was reached with a zero-inflated negative binomial GLMM formulation, using a natural logarithm link function.
	
	Schematically, the fitted model has the following structure. Being $y_{id}$ the all-cause mortality count in the $i$-$th$ ABS on day $d$, $f_{\text{NB}}$ the probability function of the associated negative binomial model NB$(\mu_{id},\theta)$, with $\mu_{id}$ and $\theta$ denoting the corresponding mean and dispersion parameters respectively, and $\pi_{id}$ the probability of an excess zero; the data generation process is given by
	
	\begin{equation}
		y_{id} \sim 
		\begin{cases}
			0 & \text{with probability } \pi_{id} + (1 - \pi_{id}) \cdot f_{\text{NB}}(0 \mid \mu_{id}, \theta) \\
			\text{NB}(\mu_{id}, \theta) & \text{with probability } (1 - \pi_{id}) \cdot f_{\text{NB}}(y_{id} \mid y_{id} > 0; \mu_{id}, \theta).
		\end{cases}
	\end{equation}
	
	The relationship with the explanatory terms is established through the (natural) logarithm link function as the linear predictor
	
	\begin{equation}
		\label{e_illustration}
		\begin{array}{l}
			{\log}\left( {\mu_{id}} \right) = {\beta _0} + {\beta _1}{\log}\left( {{x_{1id-4}}} \right) + {\beta_2}{\log}\left( {{x_{2id-4}}} \right) +  \cdots  + {\beta _5}{\log}\left( {{x_{5id-4}}} \right) + {\beta ^{(z)}}{z_{id-4}} + \\
			+ \beta _1^{(i)}{z_{id-4}}{\log}\left( {{x_{1id-4}}} \right) + \beta _2^{(i)}{z_{id-4}}{\log}\left( {{x_{2id-4}}} \right) +  \cdots  + \beta _5^{(i)}{z_{id-4}}{\log}\left( {{x_{5id-4}}} \right) + \\
			+ {\beta ^{(t)}}{t_{id-4}} + {\beta ^{(t,i)}}{t_{id-4}}{z_{id-4}} +
			{\beta ^{(p)}}{\log}\left( {p_i} \right)+
			{\eta _i} + S\left( {AB{S_i}} \right) +  \tau {s_{id}}, \\
			\mbox{ }\\
			{\rm{with}}\;\sum\limits_{j = 1}^5 {{\beta _j}}  = 0 \mbox{ and } \sum\limits_{j = 1}^5 {\beta _j^{(i)}}  = 0.
		\end{array}
	\end{equation}
	
	Regarding the fixed effects, the variables $x_{1id-4}, x_{2id-4}, x_{3id-4}, x_{4id-4}$ and $x_{5id-4}$ refer, respectively, to the daily average concentrations of the following air pollutants in the $i$-$th$ ABS on day $d-4$: particulate matter (PM\textsubscript{10}), nitrogen dioxide (NO\textsubscript{2}), ozone (O\textsubscript{3}), carbon monoxide (CO) and sulphur dioxide (SO\textsubscript{2}). These compositional variables were log-transformed using the natural logarithm, which allows the model parameters to be interpreted as elasticities, as discussed in previous sections. Moreover, the variable $z_{id-4}$ is a dummy variable indicating the daily extreme maximum temperature. It takes value 1, when the maximum temperature (or its estimated value if it was originally absent) exceeds the trigger threshold in the $i$-$th$ ABS on day $d-4$, and value 0 otherwise. Such a trigger temperature is defined as the 95-$th$ percentile of the predicted maximum temperature in the ABS during the Summer months of 2022 (June to September). Out of the 122 days under study, 15.7\% days (average across all ABS) exhibited extreme temperatures. Note that all these variables lagged by 4 days relative to the response variable. This was set after testing several lags and achieving the lowest WAIC for a lag equal to 4. Finally, the $p_i$ is an offset term accounting for the total population in the $i$-$th$ ABS.
	
	As to the random effects, the $\eta_i$ term represents an unstructured random effect accounting for heterogeneity in the ABS, e.g. due to unobserved confounders specific to the ABS and assumed invariant in time. The $S()$ term represents a structured random effect accounting for the spatial dependence among neighbouring ABSs. It is assumed to be normally distributed with zero mean and a Mat\'ern covariance function, where the covariance only depends on the distance between the ABSs involved \citep{lindgren2011explicit,saez2022spatial}. The $\tau s_{id}$ term contains structured random effects accounting for seasonality over time. Within this framework, the random effects are used as smoothers to model non-linear dependency on covariates in the linear predictor \eqref{e_illustration}.
	
	\subsection{Results}
	
	The model was implemented and fitted using the R-INLA package. As noted above, the soft constraint \eqref{e_softconstraint} was applied through the novel \texttt{A.local} function. To estimate the model parameters, we relied on a class of prior distributions that penalises complexity, called PC priors  \citep{simpson2017penalising}. A reproducible R script implementing the analysis can be found in the accompanying Appendix \ref{Appendix}.
	
	Table \ref{tab1} summarises the model coefficient estimates in terms of posterior means and posterior probabilities of being positive or negative. Note that we considered those effects with an associated posterior probability higher than 90\% for interpretation.
	
	\begin{table}[h]
		\caption{Results from the INLA estimation of the specified zero-inflated negative binomial generalised mixed model}
		\label{tab1}
		\begin{tabular}{lrr}
			\toprule
			& Mean\footnotemark[1] & Prob.\footnotemark[2] \\ 
			\midrule
			$\beta _1$ &	-0.002	&	0.623	\\
			$\beta _2$ &	0.010	&	0.914	\\
			$\beta _3$ &	-0.001	&	0.558	\\
			$\beta _4$ &	-0.006	&	0.856	\\
			$\beta _5$ &	-0.001	&	0.561	\\
			$\beta ^{(t)}$ &	0.006	&	0.983	\\
			$\beta ^{(z)}$  &	0.071	&	$>$0.99	\\
			$\beta _1^{(i)}$ &	0.025	&	0.937	\\
			$\beta _2^{(i)}$ &	0.005	&	0.613	\\
			$\beta _3^{(i)}$ &	-0.011	&	0.766	\\
			$\beta _4^{(i)}$ &	0.003	&	0.580	\\
			$\beta _5^{(i)}$ &	-0.022	&	0.922	\\
			$\beta ^{(t,i)}$ &	0.002	&	0.629	\\
			\botrule
		\end{tabular}
		\footnotetext[1]{Posterior means of the relevant model coefficients.}
		\footnotetext[2]{Posterior probabilities of being positive or negative.}
	\end{table}
	
	In the following, we provide detailed practical interpretations of these results:
	
	\begin{itemize}
		\item The coefficient $\beta _2$ has a posterior probability of being positive equal to 0.914. Then, if the ratio of the concentration of NO\textsubscript{2} to any other pollutant increases by 1\%, while keeping the total pollution constant, the expected number of deaths four days later will increase by 0.010\%. Note that this may equally apply to days of extremely high temperature or days of normal temperature, given that the corresponding interaction coefficient $\beta _2^{(i)}$ has a low posterior probability.
		
		\item The coefficient $\beta ^{(t)}$ has a posterior probability equal to 0.983 of being positive. Then, if the total pollution increases by $1/D=1/5$\%, while the relative composition remains unaltered, the expected number of deaths four days later will increase by 0.006\%. If we scale the results for an easier interpretation, when the total pollution increases by 1\%,  the expected number of deaths four days later will increase by $0.006\times5 = 0.030$\%. Note that this may equally be applied for days of extremely high temperature and days of normal temperature, since the corresponding interaction coefficient $\beta ^{(t,i)}$ has a low posterior probability of 0.629.
		
		\item The coefficient $\beta ^{(z)}$ is estimated to have a posterior probability greater than 0.999 of being positive. This indicates that the expected number of deaths will increase by 7.4\% ($e^{0.071}=1.074$) four days after the occurrence of an extreme temperature event when pollutant concentrations are on their average.
		
		\item The coefficient $\beta _1^{(i)}$ has a posterior probability of being positive equal to 0.937. Then, if the ratio of the concentration of PM\textsubscript{10} to any other pollutant increases by 1\%, while keeping the total pollution constant, the expected number of deaths will increase by 0.025\%. Note that this may only hold during days of extremely high temperature, since the corresponding main effect coefficient $\beta _1$ has a low posterior probability of 0.623. Moreover, the negative sign and high absolute value of $\beta _5^{(i)}$ (posterior probability of being negative equal to 0.922) indicates that increased mortality is expected mainly when PM\textsubscript{10} is traded for SO\textsubscript{2}.
		
	\end{itemize}
	
	\section{Final remarks}\label{s_disc}
	
	Bayesian compositional regression modelling had already been successfully addressed through INLA inference for the case of compositions in a response variable role and without considering the total \citep{adin2024automatic, figueira2024unveiling, martinez2023integrated, martinez2024flexible}. The novelty of our approach lies in handling the case of a composition in an explanatory role, for which only the less computationally efficient MCMC estimation method had been used so far, and also in considering the information provided by the total. This has implied the novel implementation of the soft constraint used in MCMC into the INLA framework. We have also extended the modelling to include moderation effects, using a parametrisation which enhances interpretability. Thus, when the response variable is log-transformed, or a logarithmic link function is used in GLM(M)s, such interpretation can be made in terms of elasticities. Notably, the proposed approach does not rely on any particular log-ratio coordinates. This, hence, avoids discussions about the convenience of using one or another representation or the lack of equivalence between them.
	
	Two key advantages of our proposal in relation to more conventional approaches used in environmental epidemiology are worth highlighting. Firstly,  if the entire composition of pollution concentrations, commonly expressed as percentages adding up to 100 or similar, is included in an ordinary model, it will suffer from perfect multicollinearity and, hence, routine parameter estimation would not be feasible. Secondly, ordinary modelling does not allow estimating the effect of the total on the response variable, since combining the composition, regardless of whether it is closed or not, with the total in the same model leads again to perfect collinearity. Moreover, our method allows us to interpret the estimated coefficients associated to individual pollutants as exchanges. If total pollution is included as a covariate in the model, then keeping it constant makes the interpretation of any change in the concentration of a pollutant to be balanced by an equivalent and opposite change in at least one other pollutant. This generally makes sense  in air pollution studies, as certain known geographical and climatological factors tend to increase some pollutants at the expense of reducing others \citep[see][and references therein] {mota2022compositional}.
	
	The method has been illustrated through an investigation of the moderation effect of air pollution and extreme temperatures on all-cause mortality. The proposed modelling brings in simple interpretations based on elasticities. Such elasticities refer to percentage increases in death counts in response to percentage increases in overall pollution or to percentage increases in the ratio of concentrations of one pollutant (NO\textsubscript{2}) to the rest. Additionally, these elasticities may be modified by the interacting variable. Some pollutant(s) (PM\textsubscript{10}) may affect mortality only, or mainly, when temperatures are extreme.
	
	Note that the approach remains valid and applicable even when the total is not considered. In such cases, the $t$ and $tz$ terms can simply be omitted from the model equation. While total levels are often of primary interest in contexts such as air, soil, and water pollution, this is not always the case in other fields. For example, in physical activity and time-use epidemiology, the total time spent in various movement behaviours each day is fixed at 24 hours (or 1440 minutes) for all individuals, rendering the total inherently irrelevant \citep{dumuid2020compositional,kuzik2025systematic}.
	
	
	
	Finally, note that the approach presented here includes only moderation effects. Related topics such as non-linear (e.g. quadratic) effects \citep{aitchison1984log} or mediation effects \citep{sohn2019compositional} are possible avenues for further research, particularly within a Bayesian framework. 
	
	\bmhead{Acknowledgements}
	
	The authors wish to thank H\aa vard Rue for his advice in implementing and programming the new feature \texttt{A.local} that allowed the soft constraint to be realized in R INLA.
	
	\section*{Declarations}
	
	\subsection*{Funding statement}  
	
	This research was supported by the Spanish Ministry of Science and Innovation and ERDF-a way of making Europe [grant number PID2021-123833OB-I00], the Spanish Ministry of Health [grant number CIBERCB06/02/1002], the Department of Research and Universities of Generalitat de Catalunya [grant numbers 2021SGR01197 and 2023-CLIMA-00037] and AGAUR and the Department of Climate Action, Food and Rural Agenda of Generalitat de Catalunya [grant number 2023-CLIMA-00037]. The funders had no role in the research process from study design to submission.
	
	\subsection*{Data availability statement}
	
	In accordance with Article 2 of Regulation 223/2009 of the Council and the European Parliament on European Statistics; Articles 13, 17.3, and 17.4 of the Spanish Law on the Public Statistical Function; Article 25.1 of Spanish Organic Law 3/2018, of December 5, on the Protection of Personal Data and Guarantee of Digital Rights; and Regulation 2016/679 of the European Union, the mortality databases of the National Institute of Statistics (INE) are subject to Statistical Confidentiality; therefore, there are restrictions on their transfer to third parties, and data are not publicly available. The datasets (properly anonymized) used and analyzed during the current study are available from the corresponding author upon reasonable request. For the rest of variables we used freely available open data.
	
	\begin{itemize}
		\item Meteorogical variables: \href{https://analisi.transparenciacatalunya.cat/ca/Medi-Ambient/Dades-meteorol-giques-de-la-XEMA/nzvn-apee/about_data}{https:// analisi.transparenciacatalunya.cat/ ca/ Medi-Ambient/ Dades-meteorol-giques-de-la-XEMA/ nzvn-apee/ about\_data}
		\item Air pollutants: \href{https://analisi.transparenciacatalunya.cat/Medi-Ambient/Qualitat-de-l-aire-als-punts-de-mesurament-autom-t/tasf-thgu/about_data}{https:// analisi.transparenciacatalunya.cat/ Medi- Ambient/ Qualitat-de-l-aire-als-punts-de-mesurament-autom-t/ tasf-thgu/ about\_data}
		\item Population data. Continuous Register Statistics: \href{https://www.ine.es/dyngs/INEbase/en/operacion.htm?c=Estadistica\_C\&cid=1254736177012\&menu=resultados\&secc=1254736195461\&idp=1254734710990\#!tabs-1254736195557}{https:// www.ine.es/ dyngs/ INEbase/ en/ operacion.htm?c=Estadistica\_C \&cid=1254736177012 \&menu=resultados \&secc= 1254736195461 \&idp=1254734710990 \#!tabs-1254736195557}
	\end{itemize}

	\subsection*{Competing interests}
	
	The authors declare no competing interests. The manuscript is an original contribution that has not been published before, whole or in part, in any format. The preprint version has been deposited in \emph{arXiv}. All authors will disclose any actual or potential conflicts of interest including any financial, personal, or other relationships with other people or organizations that could inappropriately influence or be perceived to influence their work.
	
	\subsection*{Declaration of generative AI and AI-assisted technologies}
	
	During the writing of the article the authors have not used any type of AI and AI-assisted technologies.
	
	\subsection*{Authors' Contributions}
	
	Conceptualization GC, MS; Data curation MS, MAB; Formal analysis MS, MAB; Funding acquisition MAB; Methodology GC, JPA, MS; Project administration MAB; Supervision MS, JPA; Writing -- original draft all authors; Writing -- review \& editing all authors.

	\begin{appendices}
		
		\section{Reproducible R script implementing the model fitting pipeline}\label{Appendix}
		
		{\footnotesize
			
			\noindent{\texttt{\# Load the data, the necessary packages and options}}
			
			\noindent{\noindent{\texttt{library(INLA)}}}
			
			\noindent{\texttt{inla.setOption(inla.mode="experimental")}}
			
			\noindent{\texttt{library(haven)}}
			
			\noindent{\texttt{load("file.RData")}}
			
			\noindent{\texttt{attach(file)}}
			
			\noindent{\texttt{options(contrasts=c("contr.poly","contr.treatment"))}}
			
			\noindent{\texttt{\# Define the mesh for the spatial covariance structure}}
			
			\noindent{\texttt{mesh=inla.mesh.2d(cbind(UTM\_x/1000,UTM\_y/1000),}}
			
			\noindent{\texttt{max.edge=c(15,40),}}
			
			\noindent{\texttt{offset=c(15,40), cutoff=0.5)}}
			
			\noindent{\texttt{spde=inla.spde2.pcmatern(mesh, constr=TRUE, }}
			
			\noindent{\texttt{prior.range=c(0.01,0.01), prior.sigma=c(100,0.1))}}
			
			\noindent{\texttt{field=mesh\$idx\$loc}}
			
			\noindent{\texttt{\# Soft constraint,}}
			
			\noindent{\texttt{\# main effects and centering}}
			
			\noindent{\texttt{lag\_total\_4=lag\_PM10\_4*lag\_NO2\_4*lag\_O3\_4*lag\_CO\_4*lag\_SO2\_4}}
			
			\noindent{\texttt{log\_lag\_total\_4=log(lag\_total\_4)-mean(log(lag\_total\_4),na.rm=T)}}
			
			\noindent{\texttt{X=as.matrix(cbind(}}
			
			\noindent{\texttt{log(lag\_PM10\_4)-mean(log(lag\_PM10\_4),na.rm=T), }}
			
			\noindent{\texttt{log(lag\_NO2\_4)-mean(log(lag\_NO2\_4),na.rm=T),}}
			
			\noindent{\texttt{log(lag\_O3\_4)-mean(log(lag\_O3\_4),na.rm=T),}}
			
			\noindent{\texttt{log(lag\_CO\_4)-mean(log(lag\_CO\_4),na.rm=T),} }
			
			\noindent{\texttt{log(lag\_SO2\_4)-mean(log(lag\_SO2\_4),na.rm=T)))}}
			
			\noindent{\texttt{\# Interaction and centering}}
			
			\noindent{\texttt{X\_interaction=as.matrix(cbind(}}
			
			\noindent{\texttt{(log(lag\_PM10\_4)-mean(log(lag\_PM10\_4),na.rm=T))*lag\_extreme\_heat\_4,}}
			
			\noindent{\texttt{(log(lag\_NO2\_4)-mean(log(lag\_NO2\_4),na.rm=T))*lag\_extreme\_heat\_4,}}
			
			\noindent{\texttt{(log(lag\_O3\_4)-mean(log(lag\_O3\_4),na.rm=T))*lag\_extreme\_heat\_4,}}
			
			\noindent{\texttt{(log(lag\_CO\_4)-mean(log(lag\_CO\_4),na.rm=T))*lag\_extreme\_heat\_4,}}
			
			\noindent{\texttt{(log(lag\_SO2\_4)-mean(log(lag\_SO2\_4),na.rm=T))*lag\_extreme\_heat\_4))}}
			
			\noindent{\texttt{idx=rep(NA, dim(X)[1])}}
			
			\noindent{\texttt{idx\_interaction=idx}}
			
			\noindent{\texttt{\# constr=TRUE sets the sum of coefficients in a submatrix}}
			
			\noindent{\texttt{\# of predictors to zero.}}
			
			\noindent{\texttt{formula=dead\_count \~{} 1 + }}
			
			\noindent{\texttt{f(idx, model = "iid",}}
			
			\noindent{\texttt{hyper=list(prec=list(prior="gaussian",param=c(0,10000))),}}
			
			\noindent{\texttt{constr=TRUE, }}
			
			\noindent{\texttt{A.local=X,}}
			
			\noindent{\texttt{values=1:dim(X)[2]) +}}
			
			\noindent{\texttt{f(idx\_interaction, model="iid",}}
			
			\noindent{\texttt{hyper=list(prec=list(prior="gaussian",param=c(0,10000))),}}
			
			\noindent{\texttt{constr=TRUE, }}
			
			\noindent{\texttt{A.local=X\_interaction,}}
			
			\noindent{\texttt{values=1:dim(X\_interaction)[2]) +}}
			
			\noindent{\texttt{log\_lag\_total\_4*lag\_extreme\_heat\_4 +}}
			
			\noindent{\texttt{f(field, model=spde) +}}
			
			\noindent{\texttt{f(month, model="rw1", scale.model=T, }}
			
			\noindent{\texttt{hyper=list(theta = list(prior="pc.prec", param=c(0.5,0.01))))}}
			
			\noindent{\texttt{+ offset(log(Poblacion))}}
			
			\noindent{\texttt{ptm <- proc.time()}}
			
			\noindent{\texttt{result=inla(formula, data=file, family="zeroinflatednbinomial1",}}
			
			\noindent{\texttt{control.compute=list(dic=T,waic=T), }}
			
			\noindent{\texttt{control.predictor=list(link=1,compute=TRUE),}}
			
			\noindent{\texttt{control.fixed=list(expand.factor.strategy='inla', prec=0.01,}}
			
			\noindent{\texttt{prec.intercept = 0.01))}}
			
			\noindent{\texttt{proc.time() - ptm}}
			
			\noindent{\texttt{summary(result)}
				
				\noindent{\texttt{\# Posterior probabilities}}}
			
			\noindent{\texttt{matriz=matrix(NA, length(names(result\$marginals.fixed)),1)}}
			
			\noindent{\texttt{for(i in 1: length(names(result\$marginals.fixed)))\{}}
			
			\noindent{\texttt{matriz[i,]=ifelse(result\$summary.fixed[i,1]>0,1-}}
			
			\noindent{\texttt{inla.pmarginal(0,result\$marginals.fixed[[i]]),}}
			
			\noindent{\texttt{inla.pmarginal(0,result\$marginals.fixed[[i]]))\}}}
			
			\noindent{\texttt{prob.coef.dif.zero=matriz[,1]}}
			
			\noindent{\texttt{prob.coef.dif.zero}}
			
			\noindent{\texttt{\# Goodness of fit}}
			
			\noindent{\texttt{result\$dic\$dic}}
			
			\noindent{\texttt{result\$waic\$waic}}
			
		}

	\end{appendices}

	\bibliography{clima}


\begin{thebibliography}{56}
\ifx \bisbn   \undefined \def \bisbn  #1{ISBN #1}\fi
\ifx \binits  \undefined \def \binits#1{#1}\fi
\ifx \bauthor  \undefined \def \bauthor#1{#1}\fi
\ifx \batitle  \undefined \def \batitle#1{#1}\fi
\ifx \bjtitle  \undefined \def \bjtitle#1{#1}\fi
\ifx \bvolume  \undefined \def \bvolume#1{\textbf{#1}}\fi
\ifx \byear  \undefined \def \byear#1{#1}\fi
\ifx \bissue  \undefined \def \bissue#1{#1}\fi
\ifx \bfpage  \undefined \def \bfpage#1{#1}\fi
\ifx \blpage  \undefined \def \blpage #1{#1}\fi
\ifx \burl  \undefined \def \burl#1{\textsf{#1}}\fi
\ifx \doiurl  \undefined \def \doiurl#1{\url{https://doi.org/#1}}\fi
\ifx \betal  \undefined \def \betal{\textit{et al.}}\fi
\ifx \binstitute  \undefined \def \binstitute#1{#1}\fi
\ifx \binstitutionaled  \undefined \def \binstitutionaled#1{#1}\fi
\ifx \bctitle  \undefined \def \bctitle#1{#1}\fi
\ifx \beditor  \undefined \def \beditor#1{#1}\fi
\ifx \bpublisher  \undefined \def \bpublisher#1{#1}\fi
\ifx \bbtitle  \undefined \def \bbtitle#1{#1}\fi
\ifx \bedition  \undefined \def \bedition#1{#1}\fi
\ifx \bseriesno  \undefined \def \bseriesno#1{#1}\fi
\ifx \blocation  \undefined \def \blocation#1{#1}\fi
\ifx \bsertitle  \undefined \def \bsertitle#1{#1}\fi
\ifx \bsnm \undefined \def \bsnm#1{#1}\fi
\ifx \bsuffix \undefined \def \bsuffix#1{#1}\fi
\ifx \bparticle \undefined \def \bparticle#1{#1}\fi
\ifx \barticle \undefined \def \barticle#1{#1}\fi
\bibcommenthead
\ifx \bconfdate \undefined \def \bconfdate #1{#1}\fi
\ifx \botherref \undefined \def \botherref #1{#1}\fi
\ifx \url \undefined \def \url#1{\textsf{#1}}\fi
\ifx \bchapter \undefined \def \bchapter#1{#1}\fi
\ifx \bbook \undefined \def \bbook#1{#1}\fi
\ifx \bcomment \undefined \def \bcomment#1{#1}\fi
\ifx \oauthor \undefined \def \oauthor#1{#1}\fi
\ifx \citeauthoryear \undefined \def \citeauthoryear#1{#1}\fi
\ifx \endbibitem  \undefined \def \endbibitem {}\fi
\ifx \bconflocation  \undefined \def \bconflocation#1{#1}\fi
\ifx \arxivurl  \undefined \def \arxivurl#1{\textsf{#1}}\fi
\csname PreBibitemsHook\endcsname

\bibitem[\protect\citeauthoryear{Aitchison}{1982}]{Ait1982}
\begin{barticle}
\bauthor{\bsnm{Aitchison}, \binits{J.}}:
\batitle{{{T}he statistical analysis of compositional data (with discussion)}}.
\bjtitle{Journal of the Royal Statistical Society, Series B (Statistical Methodology)}
\bvolume{44}(\bissue{2}),
\bfpage{139}--\blpage{177}
(\byear{1982})
\end{barticle}
\endbibitem

\bibitem[\protect\citeauthoryear{S{\'a}nchez-Balseca and P{\'e}rez-Foguet}{2020}]{sanchez2020spatio}
\begin{barticle}
\bauthor{\bsnm{S{\'a}nchez-Balseca}, \binits{J.}},
\bauthor{\bsnm{P{\'e}rez-Foguet}, \binits{A.}}:
\batitle{Spatio-temporal air pollution modelling using a compositional approach}.
\bjtitle{Heliyon}
\bvolume{6}(\bissue{9}),
\bfpage{04794}
(\byear{2020})
\doiurl{10.1016/j.heliyon.2020.e04794}
\end{barticle}
\endbibitem

\bibitem[\protect\citeauthoryear{Coenders et~al.}{2023}]{coenders202340}
\begin{barticle}
\bauthor{\bsnm{Coenders}, \binits{G.}},
\bauthor{\bsnm{Egozcue}, \binits{J.J.}},
\bauthor{\bsnm{Fa\v{c}evicov{\'a}}, \binits{K.}},
\bauthor{\bsnm{Navarro-L{\'o}pez}, \binits{C.}},
\bauthor{\bsnm{Palarea-Albaladejo}, \binits{J.}},
\bauthor{\bsnm{Pawlowsky-Glahn}, \binits{V.}},
\bauthor{\bsnm{Tolosana-Delgado}, \binits{R.}}:
\batitle{40 years after {A}itchison's article ``the statistical analysis of compositional data'': Where we are and where we are heading}.
\bjtitle{SORT: Statistics and Operations Research Transactions}
\bvolume{47}(\bissue{2}),
\bfpage{207}--\blpage{228}
(\byear{2023})
\doiurl{10.57645/20.8080.02.6}
\end{barticle}
\endbibitem

\bibitem[\protect\citeauthoryear{Aitchison}{1986}]{aitchison1986statistical}
\begin{bbook}
\bauthor{\bsnm{Aitchison}, \binits{J.}}:
\bbtitle{The Statistical Analysis of Compositional Data. Monographs on Statistics and Applied Probability}.
\bpublisher{Chapman and Hall},
\blocation{London, UK}
(\byear{1986})
\end{bbook}
\endbibitem

\bibitem[\protect\citeauthoryear{Pawlowsky-Glahn et~al.}{2015}]{pawlowsky2015coda}
\begin{bbook}
\bauthor{\bsnm{Pawlowsky-Glahn}, \binits{V.}},
\bauthor{\bsnm{Egozcue}, \binits{J.}},
\bauthor{\bsnm{Tolosana-Delgado}, \binits{R.}}:
\bbtitle{Modeling and Analysis of Compositional Data}.
\bpublisher{John Wiley \& Sons},
\blocation{Chichester, UK}
(\byear{2015})
\end{bbook}
\endbibitem

\bibitem[\protect\citeauthoryear{Filzmoser et~al.}{2018}]{filzmoser2018applied}
\begin{bbook}
\bauthor{\bsnm{Filzmoser}, \binits{P.}},
\bauthor{\bsnm{Hron}, \binits{K.}},
\bauthor{\bsnm{Templ}, \binits{M.}}:
\bbtitle{Applied Compositional Data Analysis}.
\bpublisher{Springer},
\blocation{New York, NY}
(\byear{2018})
\end{bbook}
\endbibitem

\bibitem[\protect\citeauthoryear{Aitchison and Bacon-Shone}{1984}]{aitchison1984log}
\begin{barticle}
\bauthor{\bsnm{Aitchison}, \binits{J.}},
\bauthor{\bsnm{Bacon-Shone}, \binits{J.}}:
\batitle{Log contrast models for experiments with mixtures}.
\bjtitle{Biometrika}
\bvolume{71}(\bissue{2}),
\bfpage{323}--\blpage{330}
(\byear{1984})
\doiurl{10.2307/2336249}
\end{barticle}
\endbibitem

\bibitem[\protect\citeauthoryear{Coenders and Pawlowsky-Glahn}{2020}]{coenders2020interpretations}
\begin{barticle}
\bauthor{\bsnm{Coenders}, \binits{G.}},
\bauthor{\bsnm{Pawlowsky-Glahn}, \binits{V.}}:
\batitle{On interpretations of tests and effect sizes in regression models with a compositional predictor}.
\bjtitle{SORT: Statistics and Operations Research Transactions}
\bvolume{44}(\bissue{1}),
\bfpage{201}--\blpage{220}
(\byear{2020})
\doiurl{10.2436/20.8080.02.100}
\end{barticle}
\endbibitem

\bibitem[\protect\citeauthoryear{van~den Boogaart et~al.}{2021}]{BFHTT2021}
\begin{barticle}
\bauthor{\bsnm{Boogaart}, \binits{K.G.}},
\bauthor{\bsnm{Filzmoser}, \binits{P.}},
\bauthor{\bsnm{Hron}, \binits{K.}},
\bauthor{\bsnm{Templ}, \binits{M.}},
\bauthor{\bsnm{Tolosana-Delgado}, \binits{R.}}:
\batitle{Classical and robust regression analysis with compositional data}.
\bjtitle{Mathematical Geosciences}
\bvolume{53}(\bissue{5}),
\bfpage{823}--\blpage{858}
(\byear{2021})
\doiurl{10.1007/s11004-020-09895-w}
\end{barticle}
\endbibitem

\bibitem[\protect\citeauthoryear{Mota~Bertran et~al.}{2022}]{mota2022compositional}
\begin{barticle}
\bauthor{\bsnm{Mota~Bertran}, \binits{A.}},
\bauthor{\bsnm{Saez}, \binits{M.}},
\bauthor{\bsnm{Coenders}, \binits{G.}}:
\batitle{Compositional and {B}ayesian inference analysis of the concentrations of air pollutants in {C}atalonia, {S}pain}.
\bjtitle{Environmental Research}
\bvolume{204},
\bfpage{112388}
(\byear{2022})
\doiurl{10.1016/j.envres.2021.112388}
\end{barticle}
\endbibitem

\bibitem[\protect\citeauthoryear{Pawlowsky-Glahn et~al.}{2015}]{pawlowsky2015tools}
\begin{barticle}
\bauthor{\bsnm{Pawlowsky-Glahn}, \binits{V.}},
\bauthor{\bsnm{Egozcue}, \binits{J.J.}},
\bauthor{\bsnm{Lovell}, \binits{D.}}:
\batitle{Tools for compositional data with a total}.
\bjtitle{Statistical Modelling}
\bvolume{15}(\bissue{2}),
\bfpage{175}--\blpage{190}
(\byear{2015})
\doiurl{10.1177/1471082X14535526}
\end{barticle}
\endbibitem

\bibitem[\protect\citeauthoryear{Coenders et~al.}{2017}]{coenders2017relative}
\begin{barticle}
\bauthor{\bsnm{Coenders}, \binits{G.}},
\bauthor{\bsnm{Mart{\'\i}n-Fern{\'a}ndez}, \binits{J.A.}},
\bauthor{\bsnm{Ferrer-Rosell}, \binits{B.}}:
\batitle{When relative and absolute information matter: Compositional predictor with a total in generalized linear models}.
\bjtitle{Statistical Modelling}
\bvolume{17}(\bissue{6}),
\bfpage{494}--\blpage{512}
(\byear{2017})
\doiurl{10.1177/1471082X17710398}
\end{barticle}
\endbibitem

\bibitem[\protect\citeauthoryear{Jaccard and Turrisi}{2003}]{jaccard2003interaction}
\begin{bbook}
\bauthor{\bsnm{Jaccard}, \binits{J.}},
\bauthor{\bsnm{Turrisi}, \binits{R.}}:
\bbtitle{Interaction Effects in Multiple Regression. Quantitative Applications in the Social Sciences 72}.
\bpublisher{Sage},
\blocation{Newbury Park, CA}
(\byear{2003})
\end{bbook}
\endbibitem

\bibitem[\protect\citeauthoryear{M{\"u}ller et~al.}{2018}]{muller2018interpretation}
\begin{barticle}
\bauthor{\bsnm{M{\"u}ller}, \binits{I.}},
\bauthor{\bsnm{Hron}, \binits{K.}},
\bauthor{\bsnm{Fi{\v s}erov\'a}, \binits{E.}},
\bauthor{\bsnm{{\v S}mahaj}, \binits{J.}},
\bauthor{\bsnm{Cakirpaloglu}, \binits{P.}},
\bauthor{\bsnm{Van{\v c}\'akov\'a}, \binits{J.}}:
\batitle{Interpretation of compositional regression with application to time budget analysis}.
\bjtitle{Austrian Journal of Statistics}
\bvolume{47}(\bissue{2}),
\bfpage{3}--\blpage{19}
(\byear{2018})
\doiurl{10.17713/ajs.v47i2.652}
\end{barticle}
\endbibitem

\bibitem[\protect\citeauthoryear{G{\'o}mez-Rubio}{2020}]{gomez2020bayesian}
\begin{bbook}
\bauthor{\bsnm{G{\'o}mez-Rubio}, \binits{V.}}:
\bbtitle{Bayesian Inference with {I}{N}{L}{A}}.
\bpublisher{Chapman and Hall/CRC Press},
\blocation{Boca Raton, FL}
(\byear{2020})
\end{bbook}
\endbibitem

\bibitem[\protect\citeauthoryear{S{\'a}nchez-Balseca and P{\'e}rez-Foguet}{2021}]{sanchez2021compositional}
\begin{barticle}
\bauthor{\bsnm{S{\'a}nchez-Balseca}, \binits{J.}},
\bauthor{\bsnm{P{\'e}rez-Foguet}, \binits{A.}}:
\batitle{Compositional spatio-temporal {P}{M}2.5 modelling in wildfires}.
\bjtitle{Atmosphere}
\bvolume{12}(\bissue{10}),
\bfpage{1309}
(\byear{2021})
\doiurl{10.3390/atmos12101309}
\end{barticle}
\endbibitem

\bibitem[\protect\citeauthoryear{Le et~al.}{2025}]{le2024bayesian}
\begin{barticle}
\bauthor{\bsnm{Le}, \binits{F.}},
\bauthor{\bsnm{Stanford}, \binits{T.E.}},
\bauthor{\bsnm{Dumuid}, \binits{D.}},
\bauthor{\bsnm{Wiley}, \binits{J.F.}}:
\batitle{Bayesian multilevel compositional data analysis: Introduction, evaluation, and application.}
\bjtitle{Psychological Methods}
(\byear{2025})
\doiurl{10.48550/arXiv.2405.03985}
\end{barticle}
\endbibitem

\bibitem[\protect\citeauthoryear{Zhang et~al.}{2025}]{zhang2025bayesian}
\begin{barticle}
\bauthor{\bsnm{Zhang}, \binits{L.}},
\bauthor{\bsnm{Zhang}, \binits{X.}},
\bauthor{\bsnm{Leach}, \binits{J.M.}},
\bauthor{\bsnm{Rahman}, \binits{A.}},
\bauthor{\bsnm{Howell}, \binits{C.R.}},
\bauthor{\bsnm{Yi}, \binits{N.}}:
\batitle{Bayesian compositional generalized linear mixed models for disease prediction using microbiome data}.
\bjtitle{BMC Bioinformatics}
\bvolume{26},
\bfpage{98}
(\byear{2025})
\doiurl{10.1186/s12859-025-06114-3}
\end{barticle}
\endbibitem

\bibitem[\protect\citeauthoryear{Zhang et~al.}{2024a}]{zhang2024bayesianordinal}
\begin{barticle}
\bauthor{\bsnm{Zhang}, \binits{L.}},
\bauthor{\bsnm{Zhang}, \binits{X.}},
\bauthor{\bsnm{Leach}, \binits{J.M.}},
\bauthor{\bsnm{Rahman}, \binits{A.F.}},
\bauthor{\bsnm{Yi}, \binits{N.}}:
\batitle{Bayesian compositional models for ordinal response}.
\bjtitle{Statistical Methods in Medical Research}
\bvolume{33}(\bissue{6}),
\bfpage{1043}--\blpage{1054}
(\byear{2024})
\doiurl{10.1177/09622802241247730}
\end{barticle}
\endbibitem

\bibitem[\protect\citeauthoryear{Zhang et~al.}{2024b}]{zhang2024bayesianglm}
\begin{barticle}
\bauthor{\bsnm{Zhang}, \binits{L.}},
\bauthor{\bsnm{Zhang}, \binits{X.}},
\bauthor{\bsnm{Yi}, \binits{N.}}:
\batitle{Bayesian compositional generalized linear models for analyzing microbiome data}.
\bjtitle{Statistics in Medicine}
\bvolume{43}(\bissue{1}),
\bfpage{141}--\blpage{155}
(\byear{2024})
\doiurl{10.1002/sim.9946}
\end{barticle}
\endbibitem

\bibitem[\protect\citeauthoryear{Scott et~al.}{2023}]{scott2023bayesian}
\begin{barticle}
\bauthor{\bsnm{Scott}, \binits{D.A.}},
\bauthor{\bsnm{Benavente}, \binits{E.}},
\bauthor{\bsnm{Libiseller-Egger}, \binits{J.}},
\bauthor{\bsnm{Fedorov}, \binits{D.}},
\bauthor{\bsnm{Phelan}, \binits{J.}},
\bauthor{\bsnm{Ilina}, \binits{E.}},
\bauthor{\bsnm{Tikhonova}, \binits{P.}},
\bauthor{\bsnm{Kudryavstev}, \binits{A.}},
\bauthor{\bsnm{Galeeva}, \binits{J.}},
\bauthor{\bsnm{Clark}, \binits{T.}},
\bauthor{\bsnm{Lewin}, \binits{A.}}:
\batitle{Bayesian compositional regression with microbiome features via variational inference}.
\bjtitle{BMC Bioinformatics}
\bvolume{24}(\bissue{1}),
\bfpage{210}
(\byear{2023})
\doiurl{10.1186/s12859-023-05219-x}
\end{barticle}
\endbibitem

\bibitem[\protect\citeauthoryear{{R Core Team}}{2025}]{r2025r}
\begin{botherref}
\oauthor{\bsnm{{R Core Team}}}:
R: A language and environment for statistical computing. {R} Foundation for Statistical Computing, {V}ienna, {AT}
(2025).
\url{http://www. R-project. org/}
Accessed 2025-04-24
\end{botherref}
\endbibitem

\bibitem[\protect\citeauthoryear{{R INLA project}}{2025}]{inlaproject2023}
\begin{botherref}
\oauthor{\bsnm{{R INLA project}}}:
{R INLA project}
(2025).
\url{http://www.r-inla.org/home}
Accessed 2025-04-24
\end{botherref}
\endbibitem

\bibitem[\protect\citeauthoryear{Mota~Bertran et~al.}{2024}]{mota2024airpollution}
\begin{barticle}
\bauthor{\bsnm{Mota~Bertran}, \binits{A.}},
\bauthor{\bsnm{Coenders}, \binits{G.}},
\bauthor{\bsnm{Plaja}, \binits{P.}},
\bauthor{\bsnm{Saez}, \binits{M.}},
\bauthor{\bsnm{Barcel{\'o}}, \binits{M.A.}}:
\batitle{Air pollution and children's mental health in rural areas. {C}ompositional spatio-temporal model}.
\bjtitle{Scientific Reports}
\bvolume{14},
\bfpage{19363}
(\byear{2024})
\doiurl{10.1038/s41598-024-70024-2}
\end{barticle}
\endbibitem

\bibitem[\protect\citeauthoryear{Dargel and Thomas-Agnan}{2024}]{dargel2024link}
\begin{barticle}
\bauthor{\bsnm{Dargel}, \binits{L.}},
\bauthor{\bsnm{Thomas-Agnan}, \binits{C.}}:
\batitle{The link between multiplicative competitive interaction models and compositional data regression with a total}.
\bjtitle{Journal of Applied Statistics}
\bvolume{51}(\bissue{14}),
\bfpage{2929}--\blpage{2960}
(\byear{2024})
\doiurl{10.1080/02664763.2024.2329923}
\end{barticle}
\endbibitem

\bibitem[\protect\citeauthoryear{Morais and Thomas-Agnan}{2021}]{morais2021impact}
\begin{barticle}
\bauthor{\bsnm{Morais}, \binits{J.}},
\bauthor{\bsnm{Thomas-Agnan}, \binits{C.}}:
\batitle{Impact of covariates in compositional models and simplicial derivatives}.
\bjtitle{Austrian Journal of Statistics}
\bvolume{50}(\bissue{2}),
\bfpage{1}--\blpage{15}
(\byear{2021})
\doiurl{10.17713/ajs.v50i2.1069}
\end{barticle}
\endbibitem

\bibitem[\protect\citeauthoryear{S{\'a}nchez-Balseca and P{\'e}rez-Foguet}{2022}]{sanchez2022spatially}
\begin{barticle}
\bauthor{\bsnm{S{\'a}nchez-Balseca}, \binits{J.}},
\bauthor{\bsnm{P{\'e}rez-Foguet}, \binits{A.}}:
\batitle{Spatially-structured human mortality modelling using air pollutants with a compositional approach}.
\bjtitle{Science of The Total Environment}
\bvolume{813},
\bfpage{152486}
(\byear{2022})
\doiurl{10.1016/j.scitotenv.2021.152486}
\end{barticle}
\endbibitem

\bibitem[\protect\citeauthoryear{Alari et~al.}{2023}]{alari2023role}
\begin{barticle}
\bauthor{\bsnm{Alari}, \binits{A.}},
\bauthor{\bsnm{Chen}, \binits{C.}},
\bauthor{\bsnm{Schwarz}, \binits{L.}},
\bauthor{\bsnm{Hdansen}, \binits{K.}},
\bauthor{\bsnm{Chaix}, \binits{B.}},
\bauthor{\bsnm{Benmarhnia}, \binits{T.}}:
\batitle{The role of ozone as a mediator of the relationship between heat waves and mortality in 15 {F}rench urban areas}.
\bjtitle{American Journal of Epidemiology}
\bvolume{192}(\bissue{6}),
\bfpage{949}--\blpage{962}
(\byear{2023})
\doiurl{10.1093/aje/kwad032}
\end{barticle}
\endbibitem

\bibitem[\protect\citeauthoryear{Analitis et~al.}{2014}]{analitis2014effects}
\begin{barticle}
\bauthor{\bsnm{Analitis}, \binits{A.}},
\bauthor{\bsnm{Michelozzi}, \binits{P.}},
\bauthor{\bsnm{D'Ippoliti}, \binits{D.}},
\bauthor{\bsnm{de'Donato}, \binits{F.}},
\bauthor{\bsnm{Menne}, \binits{B.}},
\bauthor{\bsnm{Matthies}, \binits{F.}},
\bauthor{\bsnm{Atkinson}, \binits{R.W.}},
\bauthor{\bsnm{I{\~n}iguez}, \binits{C.}},
\bauthor{\bsnm{Basaga{\~n}a}, \binits{X.}},
\bauthor{\bsnm{Schneider}, \binits{A.}},
\bauthor{\bsnm{Lefranc}, \binits{A.}},
\bauthor{\bsnm{Paldy}, \binits{A.}},
\bauthor{\bsnm{Bisanti}, \binits{L.}},
\bauthor{\bsnm{Katsouyanni}, \binits{K.}}:
\batitle{Effects of heat waves on mortality: Effect modification and confounding by air pollutants}.
\bjtitle{Epidemiology}
\bvolume{25}(\bissue{1}),
\bfpage{15}--\blpage{22}
(\byear{2014})
\doiurl{10.1097/EDE.0b013e31828ac01b}
\end{barticle}
\endbibitem

\bibitem[\protect\citeauthoryear{Dear et~al.}{2005}]{dear2005effects}
\begin{barticle}
\bauthor{\bsnm{Dear}, \binits{K.}},
\bauthor{\bsnm{Ranmuthugala}, \binits{G.}},
\bauthor{\bsnm{Kjellstr{\"o}m}, \binits{T.}},
\bauthor{\bsnm{Skinner}, \binits{C.}},
\bauthor{\bsnm{Hanigan}, \binits{I.}}:
\batitle{Effects of temperature and ozone on daily mortality during the august 2003 heat wave in {F}rance}.
\bjtitle{Archives of Environmental \& Occupational Health}
\bvolume{60}(\bissue{4}),
\bfpage{205}--\blpage{212}
(\byear{2005})
\doiurl{10.3200/AEOH.60.4.205-212}
\end{barticle}
\endbibitem

\bibitem[\protect\citeauthoryear{Du et~al.}{2024}]{du2024exposure}
\begin{barticle}
\bauthor{\bsnm{Du}, \binits{H.}},
\bauthor{\bsnm{Yan}, \binits{M.}},
\bauthor{\bsnm{Liu}, \binits{X.}},
\bauthor{\bsnm{Zhong}, \binits{Y.}},
\bauthor{\bsnm{Ban}, \binits{J.}},
\bauthor{\bsnm{Lu}, \binits{K.}},
\bauthor{\bsnm{Li}, \binits{T.}}:
\batitle{Exposure to concurrent heatwaves and ozone pollution and associations with mortality risk: A nationwide study in {C}hina}.
\bjtitle{Environmental Health Perspectives}
\bvolume{132}(\bissue{4}),
\bfpage{047012}
(\byear{2024})
\doiurl{10.1289/EHP13790}
\end{barticle}
\endbibitem

\bibitem[\protect\citeauthoryear{Qi et~al.}{2023}]{qi2023modification}
\begin{barticle}
\bauthor{\bsnm{Qi}, \binits{J.}},
\bauthor{\bsnm{Wang}, \binits{Y.}},
\bauthor{\bsnm{Wang}, \binits{L.}},
\bauthor{\bsnm{Cao}, \binits{R.}},
\bauthor{\bsnm{Huang}, \binits{J.}},
\bauthor{\bsnm{Li}, \binits{G.}},
\bauthor{\bsnm{Yin}, \binits{P.}}:
\batitle{The modification effect of ozone pollution on the associations between heat wave and cardiovascular mortality}.
\bjtitle{The Innovation Medicine}
\bvolume{1}(\bissue{10}),
\bfpage{59717}
(\byear{2023})
\doiurl{10.59717/j.xinn-med.2023.100043}
\end{barticle}
\endbibitem

\bibitem[\protect\citeauthoryear{Martenies et~al.}{2022}]{Martenies2022using}
\begin{barticle}
\bauthor{\bsnm{Martenies}, \binits{S.E.}},
\bauthor{\bsnm{Hoskovec}, \binits{L.}},
\bauthor{\bsnm{Wilson}, \binits{A.}},
\bauthor{\bsnm{Moore}, \binits{B.F.}},
\bauthor{\bsnm{Starling}, \binits{A.P.}},
\bauthor{\bsnm{Allshouse}, \binits{W.B.}},
\bauthor{\bsnm{Adgate}, \binits{J.L.}},
\bauthor{\bsnm{Dabelea}, \binits{D.}},
\bauthor{\bsnm{Magzamen}, \binits{S.}}:
\batitle{Using non-parametric {B}ayes shrinkage to assess relationships between multiple environmental and social stressors and neonatal size and body composition in the healthy start cohort}.
\bjtitle{Environmental Health}
\bvolume{21}(\bissue{1}),
\bfpage{111}
(\byear{2022})
\doiurl{10.1186/s12940-022-00934-z}
\end{barticle}
\endbibitem

\bibitem[\protect\citeauthoryear{Wang et~al.}{2020}]{Wang2020sperm}
\begin{barticle}
\bauthor{\bsnm{Wang}, \binits{X.}},
\bauthor{\bsnm{Tian}, \binits{X.}},
\bauthor{\bsnm{Ye}, \binits{B.}},
\bauthor{\bsnm{Zhang}, \binits{Y.}},
\bauthor{\bsnm{Zhang}, \binits{X.}},
\bauthor{\bsnm{Huang}, \binits{S.}},
\bauthor{\bsnm{Li}, \binits{C.}},
\bauthor{\bsnm{Wu}, \binits{S.}},
\bauthor{\bsnm{Li}, \binits{R.}},
\bauthor{\bsnm{Zou}, \binits{Y.}},
\bauthor{\bsnm{Liao}, \binits{J.}},
\bauthor{\bsnm{Yang}, \binits{J.}},
\bauthor{\bsnm{Ma}, \binits{L.}}:
\batitle{The association between ambient temperature and sperm quality in {W}uhan, {C}hina}.
\bjtitle{Environmental Health}
\bvolume{19}(\bissue{1}),
\bfpage{44}
(\byear{2020})
\doiurl{10.1186/s12940-020-00595-w}
\end{barticle}
\endbibitem

\bibitem[\protect\citeauthoryear{Landguth et~al.}{2024}]{Landguth2024seasonal}
\begin{barticle}
\bauthor{\bsnm{Landguth}, \binits{E.L.}},
\bauthor{\bsnm{Knudson}, \binits{J.}},
\bauthor{\bsnm{Graham}, \binits{J.}},
\bauthor{\bsnm{Orr}, \binits{A.}},
\bauthor{\bsnm{Coyle}, \binits{E.A.}},
\bauthor{\bsnm{Smith}, \binits{P.}},
\bauthor{\bsnm{Semmens}, \binits{E.O.}},
\bauthor{\bsnm{Noonan}, \binits{C.}}:
\batitle{Seasonal extreme temperatures and short-term fine particulate matter increases pediatric respiratory healthcare encounters in a sparsely populated region of the intermountain western {U}nited {S}tates}.
\bjtitle{Environmental Health}
\bvolume{23}(\bissue{1}),
\bfpage{40}
(\byear{2024})
\doiurl{10.1186/s12940-024-01082-2}
\end{barticle}
\endbibitem

\bibitem[\protect\citeauthoryear{Xu et~al.}{2023}]{xu2023extreme}
\begin{barticle}
\bauthor{\bsnm{Xu}, \binits{R.}},
\bauthor{\bsnm{Huang}, \binits{S.}},
\bauthor{\bsnm{Shi}, \binits{C.}},
\bauthor{\bsnm{Wang}, \binits{R.}},
\bauthor{\bsnm{Liu}, \binits{T.}},
\bauthor{\bsnm{Li}, \binits{Y.}},
\bauthor{\bsnm{Zheng}, \binits{Y.}},
\bauthor{\bsnm{Lv}, \binits{Z.}},
\bauthor{\bsnm{Wei}, \binits{J.}},
\bauthor{\bsnm{Sun}, \binits{H.}},
\bauthor{\bsnm{Liu}, \binits{Y.}}:
\batitle{Extreme temperature events, fine particulate matter, and myocardial infarction mortality}.
\bjtitle{Circulation}
\bvolume{148}(\bissue{4}),
\bfpage{312}--\blpage{323}
(\byear{2023})
\doiurl{10.1161/CIRCULATIONAHA.122.063504}
\end{barticle}
\endbibitem

\bibitem[\protect\citeauthoryear{Lindgren and Rue}{2015}]{lindgren2015bayesian}
\begin{barticle}
\bauthor{\bsnm{Lindgren}, \binits{F.}},
\bauthor{\bsnm{Rue}, \binits{H.}}:
\batitle{Bayesian spatial modelling with {R-INLA}}.
\bjtitle{Journal of Statistical Software}
\bvolume{63},
\bfpage{1}--\blpage{25}
(\byear{2015})
\end{barticle}
\endbibitem

\bibitem[\protect\citeauthoryear{Rue et~al.}{2009}]{rue2009approximate}
\begin{barticle}
\bauthor{\bsnm{Rue}, \binits{H.}},
\bauthor{\bsnm{Martino}, \binits{S.}},
\bauthor{\bsnm{Chopin}, \binits{N.}}:
\batitle{Approximate {B}ayesian inference for latent {G}aussian models by using integrated nested {L}aplace approximations}.
\bjtitle{Journal of the Royal Statistical Society Series B: Statistical Methodology}
\bvolume{71}(\bissue{2}),
\bfpage{319}--\blpage{392}
(\byear{2009})
\doiurl{j.1467-9868.2008.00700.x}
\end{barticle}
\endbibitem

\bibitem[\protect\citeauthoryear{Lindgren et~al.}{2011}]{lindgren2011explicit}
\begin{barticle}
\bauthor{\bsnm{Lindgren}, \binits{F.}},
\bauthor{\bsnm{Rue}, \binits{H.}},
\bauthor{\bsnm{Lindstr{\"o}m}, \binits{J.}}:
\batitle{An explicit link between {G}aussian fields and {G}aussian {M}arkov random fields: The stochastic partial differential equation approach}.
\bjtitle{Journal of the Royal Statistical Society Series B: Statistical Methodology}
\bvolume{73}(\bissue{4}),
\bfpage{423}--\blpage{498}
(\byear{2011})
\doiurl{10.1111/j.1467-9868.2011.00777.x}
\end{barticle}
\endbibitem

\bibitem[\protect\citeauthoryear{Cameletti et~al.}{2013}]{cameletti2013spatio}
\begin{barticle}
\bauthor{\bsnm{Cameletti}, \binits{M.}},
\bauthor{\bsnm{Lindgren}, \binits{F.}},
\bauthor{\bsnm{Simpson}, \binits{D.}},
\bauthor{\bsnm{Rue}, \binits{H.}}:
\batitle{Spatio-temporal modeling of particulate matter concentration through the {SPDE} approach}.
\bjtitle{AStA Advances in Statistical Analysis}
\bvolume{97},
\bfpage{109}--\blpage{131}
(\byear{2013})
\doiurl{10.1007/s10182-012-0196-3}
\end{barticle}
\endbibitem

\bibitem[\protect\citeauthoryear{Saez and Barcel{\'o}}{2022}]{saez2022spatial}
\begin{barticle}
\bauthor{\bsnm{Saez}, \binits{M.}},
\bauthor{\bsnm{Barcel{\'o}}, \binits{M.A.}}:
\batitle{Spatial prediction of air pollution levels using a hierarchical {B}ayesian spatiotemporal model in {C}atalonia, {S}pain}.
\bjtitle{Environmental Modelling \& Software}
\bvolume{151},
\bfpage{105369}
(\byear{2022})
\doiurl{10.1016/j.envsoft.2022.105369}
\end{barticle}
\endbibitem

\bibitem[\protect\citeauthoryear{{Agencia Espa\~{n}ola de Meteorolog\'ia -AEMET-}}{2024}]{AEMET}
\begin{botherref}
\oauthor{\bsnm{{Agencia Espa\~{n}ola de Meteorolog\'ia -AEMET-}}}:
Olas de calor en {E}spa\~{n}a desde 1975
(2024)
\end{botherref}
\endbibitem

\bibitem[\protect\citeauthoryear{METEOCAT}{2022}]{METEOCAT}
\begin{botherref}
\oauthor{\bsnm{METEOCAT}}:
Balan\c{c} d'una de les onades de calor m\'es persistents mesurades a {C}atalunya, juliol 2022
(2022)
\end{botherref}
\endbibitem

\bibitem[\protect\citeauthoryear{{Departament de Territori, Habitatge i Transici\'o Ecol\`ogica}}{2025a}]{METEOCATDades}
\begin{botherref}
\oauthor{\bsnm{{Departament de Territori, Habitatge i Transici\'o Ecol\`ogica}}}:
Dades meteorol\`ogiques de la {X}{E}{M}{A}
(2025)
\end{botherref}
\endbibitem

\bibitem[\protect\citeauthoryear{{Departament de Territori, Habitatge i Transici\'o Ecol\`ogica}}{2025b}]{XVPCA}
\begin{botherref}
\oauthor{\bsnm{{Departament de Territori, Habitatge i Transici\'o Ecol\`ogica}}}:
Dades dels punts de mesurament autom\`atics de la Xarxa de Vigil\`ancia i Previsi\'o de la Contaminaci\'o Atmosf\`erica
(2025)
\end{botherref}
\endbibitem

\bibitem[\protect\citeauthoryear{Rue et~al.}{2017}]{rue2017bayesian}
\begin{barticle}
\bauthor{\bsnm{Rue}, \binits{H.}},
\bauthor{\bsnm{Riebler}, \binits{A.}},
\bauthor{\bsnm{S{\o}rbye}, \binits{S.H.}},
\bauthor{\bsnm{Illian}, \binits{J.B.}},
\bauthor{\bsnm{Simpson}, \binits{D.P.}},
\bauthor{\bsnm{Lindgren}, \binits{F.K.}}:
\batitle{Bayesian computing with {INLA}: A review}.
\bjtitle{Annual Review of Statistics and Its Application}
\bvolume{4}(\bissue{1}),
\bfpage{395}--\blpage{421}
(\byear{2017})
\doiurl{annurev-statistics-060116-054045}
\end{barticle}
\endbibitem

\bibitem[\protect\citeauthoryear{Van~Niekerk et~al.}{2023}]{van2023new}
\begin{barticle}
\bauthor{\bsnm{Van~Niekerk}, \binits{J.}},
\bauthor{\bsnm{Krainski}, \binits{E.}},
\bauthor{\bsnm{Rustand}, \binits{D.}},
\bauthor{\bsnm{Rue}, \binits{H.}}:
\batitle{A new avenue for {B}ayesian inference with {INLA}}.
\bjtitle{Computational Statistics \& Data Analysis}
\bvolume{181},
\bfpage{107692}
(\byear{2023})
\doiurl{10.1016/j.csda.2023.107692}
\end{barticle}
\endbibitem

\bibitem[\protect\citeauthoryear{Watanabe}{2010}]{watanabe2010asymptotic}
\begin{barticle}
\bauthor{\bsnm{Watanabe}, \binits{S.}}:
\batitle{Asymptotic equivalence of {B}ayes cross validation and widely applicable information criterion in singular learning theory.}
\bjtitle{Journal of Machine Learning Research}
\bvolume{11}(\bissue{12}),
\bfpage{3571}--\blpage{3594}
(\byear{2010})
\end{barticle}
\endbibitem

\bibitem[\protect\citeauthoryear{Simpson et~al.}{2017}]{simpson2017penalising}
\begin{barticle}
\bauthor{\bsnm{Simpson}, \binits{D.}},
\bauthor{\bsnm{Rue}, \binits{H.}},
\bauthor{\bsnm{Riebler}, \binits{A.}},
\bauthor{\bsnm{Martins}, \binits{T.G.}},
\bauthor{\bsnm{S{\o}rbye}, \binits{S.H.}}:
\batitle{Penalising model component complexity: A principled, practical approach to constructing priors}.
\bjtitle{Statistical Science}
\bvolume{32}(\bissue{1}),
\bfpage{1}--\blpage{46}
(\byear{2017})
\doiurl{10.1214/16-STS576}
\end{barticle}
\endbibitem

\bibitem[\protect\citeauthoryear{Adin et~al.}{2024}]{adin2024automatic}
\begin{barticle}
\bauthor{\bsnm{Adin}, \binits{A.}},
\bauthor{\bsnm{Krainski}, \binits{E.T.}},
\bauthor{\bsnm{Lenzi}, \binits{A.}},
\bauthor{\bsnm{Liu}, \binits{Z.}},
\bauthor{\bsnm{Mart{\'\i}nez-Minaya}, \binits{J.}},
\bauthor{\bsnm{Rue}, \binits{H.}}:
\batitle{Automatic cross-validation in structured models: Is it time to leave out leave-one-out?}
\bjtitle{Spatial Statistics}
\bvolume{62},
\bfpage{100843}
(\byear{2024})
\doiurl{10.1016/j.spasta.2024.100843}
\end{barticle}
\endbibitem

\bibitem[\protect\citeauthoryear{Figueira et~al.}{2025}]{figueira2024unveiling}
\begin{barticle}
\bauthor{\bsnm{Figueira}, \binits{M.}},
\bauthor{\bsnm{Guarner}, \binits{C.}},
\bauthor{\bsnm{Conesa}, \binits{D.}},
\bauthor{\bsnm{L\'opez-Qu\'ilez}, \binits{A.}},
\bauthor{\bsnm{Krisztin}, \binits{T.}}:
\batitle{Unveiling land use dynamics: Insights from a hierarchical {B}ayesian spatio-temporal modelling of compositional data}.
\bjtitle{Journal of Agricultural, Biological, and Environmental Statistics}
(\byear{2025})
\doiurl{10.1007/s13253-025-00678-6}
\end{barticle}
\endbibitem

\bibitem[\protect\citeauthoryear{Mart{\'\i}nez-Minaya et~al.}{2023}]{martinez2023integrated}
\begin{barticle}
\bauthor{\bsnm{Mart{\'\i}nez-Minaya}, \binits{J.}},
\bauthor{\bsnm{Lindgren}, \binits{F.}},
\bauthor{\bsnm{L{\'o}pez-Qu{\'\i}lez}, \binits{A.}},
\bauthor{\bsnm{Simpson}, \binits{D.}},
\bauthor{\bsnm{Conesa}, \binits{D.}}:
\batitle{The integrated nested {L}aplace approximation for fitting {D}irichlet regression models}.
\bjtitle{Journal of Computational and Graphical Statistics}
\bvolume{32}(\bissue{3}),
\bfpage{805}--\blpage{823}
(\byear{2023})
\doiurl{10.1080/10618600.2022.2144330}
\end{barticle}
\endbibitem

\bibitem[\protect\citeauthoryear{Mart{\'\i}nez-Minaya and Rue}{2024}]{martinez2024flexible}
\begin{barticle}
\bauthor{\bsnm{Mart{\'\i}nez-Minaya}, \binits{J.}},
\bauthor{\bsnm{Rue}, \binits{H.}}:
\batitle{A flexible {B}ayesian tool for {C}o{D}a mixed models: Logistic-normal distribution with {D}irichlet covariance}.
\bjtitle{Statistics and Computing}
\bvolume{34}(\bissue{3}),
\bfpage{116}
(\byear{2024})
\doiurl{10.1007/s11222-024-10427-3}
\end{barticle}
\endbibitem

\bibitem[\protect\citeauthoryear{Dumuid et~al.}{2020}]{dumuid2020compositional}
\begin{barticle}
\bauthor{\bsnm{Dumuid}, \binits{D.}},
\bauthor{\bsnm{Pedi{\v{s}}i{\'c}}, \binits{{\v{Z}}.}},
\bauthor{\bsnm{Palarea-Albaladejo}, \binits{J.}},
\bauthor{\bsnm{Mart{\'\i}n-Fern{\'a}ndez}, \binits{J.A.}},
\bauthor{\bsnm{Hron}, \binits{K.}},
\bauthor{\bsnm{Olds}, \binits{T.}}:
\batitle{Compositional data analysis in time-use epidemiology: what, why, how}.
\bjtitle{International Journal of Environmental Research and Public Health}
\bvolume{17}(\bissue{7}),
\bfpage{2220}
(\byear{2020})
\doiurl{10.3390/ijerph17072220}
\end{barticle}
\endbibitem

\bibitem[\protect\citeauthoryear{Kuzik et~al.}{2025}]{kuzik2025systematic}
\begin{barticle}
\bauthor{\bsnm{Kuzik}, \binits{N.}},
\bauthor{\bsnm{Duncan}, \binits{M.J.}},
\bauthor{\bsnm{Beshara}, \binits{N.}},
\bauthor{\bsnm{MacDonald}, \binits{M.}},
\bauthor{\bsnm{Silva}, \binits{D.A.S.}},
\bauthor{\bsnm{Tremblay}, \binits{M.S.}}:
\batitle{A systematic review and meta-analysis of the first decade of compositional data analyses of 24-hour movement behaviours, health, and well-being in school-aged children}.
\bjtitle{Journal of Activity, Sedentary and Sleep Behaviors}
\bvolume{4}(\bissue{1}),
\bfpage{4}
(\byear{2025})
\doiurl{10.1186/s44167-025-00076-w}
\end{barticle}
\endbibitem

\bibitem[\protect\citeauthoryear{Sohn and Li}{2019}]{sohn2019compositional}
\begin{barticle}
\bauthor{\bsnm{Sohn}, \binits{M.B.}},
\bauthor{\bsnm{Li}, \binits{H.}}:
\batitle{Compositional mediation analysis for microbiome studies}.
\bjtitle{The Annals of Applied Statistics}
\bvolume{13}(\bissue{1}),
\bfpage{661}--\blpage{681}
(\byear{2019})
\doiurl{10.1214/18-AOAS1210}
\end{barticle}
\endbibitem

\end{thebibliography}
	
\end{document}